\PassOptionsToPackage{dvipsnames}{xcolor}
\documentclass[sn-basic,iicol]{sn-jnl}% Default with double column layout

\jyear{2023}%

\theoremstyle{thmstyleone}%
\theoremstyle{thmstyletwo}%

\theoremstyle{thmstylethree}%

\begin{document}

\title[Article Title]{Topology Optimization with Text-Guided Stylization}

\author*[1]{\fnm{Shengze} \sur{Zhong}}\email{zhong@sens.sys.es.osaka-u.ac.jp}

\author[1]{\fnm{Parinya} \sur{Punpongsanon}}

\author[1]{\fnm{Daisuke} \sur{Iwai}}

\author[1]{\fnm{Kosuke} \sur{Sato}}

\affil[1]{\orgdiv{Graduate School of Engineering Science}, \orgname{Osaka University}, \orgaddress{\state{Osaka}, \country{Japan}}}

\abstract{We propose an approach for the generation of topology-optimized structures with text-guided appearance stylization. This methodology aims to enrich the concurrent design of a structure's physical functionality and aesthetic appearance. Users can effortlessly input descriptive text to govern the style of the structure. Our system employs a hash-encoded neural network as the implicit structure representation backbone,  which serves as the foundation for the co-optimization of structural mechanical performance, style, and connectivity, to ensure full-color, high-quality 3D-printable solutions. We substantiate the effectiveness of our system through extensive comparisons, demonstrations, and a 3D printing test.}


\maketitle

\section{Introduction}
% Revised as: \textcolor{red}{revise}
Topology optimization is a mathematical method that automatically designs structures with optimal performance under physical boundary conditions and constraints \citep{rozvany2009critical}. Its utilization has progressively extended into engineering domains like automotive engineering \citep{yang1995automotive} and aerospace engineering \citep{zhu2016topology, aage2017giga}, where the demand for high-performance structures is paramount. With the advancement of topology optimization algorithms, users can now generate high-quality structures by manipulating a small number of parameters, such as material properties and manufacturing costs \citep{kazi2017dreamsketch, chen2018forte, ma2021creating}. Consequently, even individuals with limited expertise can effectively wield optimization tools following brief training \citep{nobel2016improving}. This simplicity of structure design and reduced demand for user expertise have facilitated the wide-ranging application of topology optimization across domains, including garment design \citep{zhang2019customization} and the development of musical instruments \citep{yu2013topology, li2016acoustic}.

The advent of commercially available topology-optimized products has sparked an upsurge in the desire for visually appealing designs. The appearance of a product holds considerable sway over user preferences, with aesthetic and symbolic characteristics assuming dominant roles \citep{creusen2005different}. However, designing structures that balance both performance and appearance remains a challenging task, even for seasoned designers. We refer to this task as stylized topology optimization.

Due to the difficulty in building a comprehensive and differentiable description of the structure style, previous investigations into stylized topology optimization have primarily relied on texture-based approaches to guide the visual aesthetics of the structure \citep{martinez2015structure, hu2019texture, navez2022topology}. These methods locally apply geometric features to the structure that align with the user-provided texture, albeit at the expense of a holistic stylization perspective. Furthermore, they necessitate the laborious task of manually designing textures, thus compromising user convenience. On the other hand, Loos et al. creatively introduced a general evaluation criterion, unity-in-variety, to assess structural style \citep{loos2022towards}. This approach has demonstrated its ability to improve user aesthetic preference while requiring manually adjusted structures.

We are motivated to address the aforementioned issues by exploring a comprehensive and differentiable structural stylization evaluation metric, for bringing enhanced stylization expressiveness. Inspired by the recent success of large-scale image-text neural networks in content generation \citep{radford2021learning}, we present a novel approach to topology optimization that incorporates text-guided stylization. By inputting the desired appearance's descriptive text alongside the physical boundary conditions, this method can generate mechanically optimized, full-color stylized, and 3D printable structures.
% =========================

Our research object can be succinctly viewed as a multi-objective optimization problem of an implicitly neural represented structure \citep{xie2022neural}. To this end, we employ a hash-encoded neural network \citep{muller2022instant} to encode coordinates into color and density, effectively capturing both the topology and appearance of the structure. This approach offers a superior representation of structural high-frequency details and demonstrates a faster convergence rate when compared to methods relying on Fourier-featured positional encoding \citep{tancik2020fourier, sitzmann2020implicit}. Subsequently, we leverage a pre-trained image-text neural network called CLIP \citep{radford2021learning} to evaluate the latent feature similarity between the appearance image of the structure and the user's textual description. This evaluation guides the stylization process, facilitating effective control over the desired appearance. Furthermore, we conduct structure connectivity optimization based on connected component labeling \citep{he2017connected} to ensure the structure can be 3D-printed in one piece. In the experiments, we analyzed the structural mechanical performance in benchmark tasks \citep{valdez2017topology}, showcased various stylized structures, presented the structure stylization control achievable with this research, and concluded with a 3D printing test.

In summary, we present a text-driven stylized topology optimization method. It employs more user-friendly text-based guidance for the appearance design of a diverse range of topology-optimized structures, and for the first time, takes into account the overall style of the structure, in full color.

\section{Related works}\label{sec2}
\subsection{Topology optimization methods}\label{subsec2.1}
In mechanical engineering, topology optimization is a method that maximizes structural performance by reallocating the spatial distribution of materials \citep{sigmund2013topology}. Its applications include a range of areas, such as enhancing the flexibility of structures \citep{bruggi2012topology}, adjusting the natural vibration frequency \citep{tsai2013structural}, and optimizing heat conduction \citep{dbouk2017review}. Based on the representation of structures, topology optimization methods can be categorized into explicit and implicit, with representative algorithms being solid isotropic material with penalization (SIMP) \citep{andreassen2011efficient}, bi-directional evolutionary structural optimization (BESO) \citep{huang2009bi}, and level-set-based methods \citep{wang2003level, zhang2016new}. Recently, the advent of implicit neural representation (INR) in topology optimization \citep{chandrasekhar2021tounn, woldseth2022use} has enabled novel functionalities, such as arbitrary resolution sampling or solution space generation \citep{zehnder2021ntopo, 10.1111:cgf.14700}. 

Regardless of the representation form of the structure, the primary flow of topology optimization involves projecting the structure onto a finite-element mesh for mechanical performance analysis and then back-propagating the gradient of mechanical performance to update the representation of the structure. This optimization process typically yields a flat solution space, wherein multiple local optima coexist for the given boundary conditions \citep{sigmund1998numerical}. Exploiting this characteristic, we are afforded an ample computational realm to introduce stylized design aspects to the structural appearance.

\subsection{Stylized topology optimization}\label{subsec2.2}
On this basis, previous works have explored various stylized topology optimization methods with texture guidance, in order to enhance the structural aesthetics. Martinez et al. \citep{martinez2015structure} proposed using exemplars as a guide to stylize 2D topology-optimized structures, endowing the structures with features resembling the exemplar. Their work entailed deriving the first derivative of the structural similarity to the exemplar and re-formulating the multi-objective optimization problem to balance mechanical performance and appearance. Subsequently, Hu et al. \citep{hu2019texture} introduced a texture-guided generative structural design method that simultaneously generates a series of stylized structures based on textures. Navez et al. \citep{navez2022topology} recently extended these efforts from 2D to 3D, with enhanced local stylization control. Additionally, Loos et al. \citep{loos2022towards} analyzed the aesthetics of topology-optimized structures using the principle of unity-in-variety in industrial design and proposed a simulation for improving the design. These studies have well demonstrated the potential and applications of stylizing topology-optimized structures. 

However, there is still no widely accepted structural style evaluation metric in the field of topology optimization, which is due to the highly abstract and complex nature of structural aesthetics and styles. As a consequence, previous investigations have concentrated on specific structural characteristics, such as local geometric patterns, in an attempt to establish quantifiable style evaluation metrics. However, these approaches often sacrifice the overall expressiveness of stylization. Moreover, highly specialized style metrics lead to the lacking of stylization controllability. Furthermore, they have not fully considered the connectivity of 3D-printed structures. Our study addresses these limitations.

\subsection{Text-guided generation}\label{subsec2.3}
The challenge of stylized topology optimization lies in establishing a differentiable and objective evaluation criterion of structural appearance. Recently, data-driven methods show a feasible solution. For instance, Chen et al. proposed a neural network that bridges shape and its human-evaluated aesthetics. After training, the network prediction of shape aesthetics could be leveraged to guide the beautification of novel input shapes \citep{chen2022learning}.

The proposed text-guided structure stylization method is further inspired by the recently prominent text-guided AI drawing and modeling \citep{frans2021clipdraw, rombach2022high, jain2022zero, poole2022dreamfusion}, where text, one of the most common and expressive mediums, is utilized to guide the stylization and creation. This technique generally relies on a neural network trained on huge datasets of image-text pairs, to create a multi-modal relationship in between. Then the network estimates the cross-modal similarity between the generated object (often rendered as an image) and the user-input text description and performs optimization. For example, Kevin et al. utilized CLIP guidance to generate drawings based on text input \citep{frans2021clipdraw}. Michel et al. proposed a Text2Mesh system that optimizes the position and color of mesh vertices through CLIP guidance, thereby generating 3D objects that conform to the textual description \citep{michel2022text2mesh}. Though previous researches show prominent simplicity and expressiveness in shape creation, those involving physical properties (e.g., mechanical performance, connectivity of 3D-printed structure) have not been fully explored.

To address this disparity, we present a novel topology optimization method enriched with text-guided stylization. Our approach enables the generation of visually captivating and structurally robust designs without the need for arduous geometric editing. It also featured much stronger convergence than prevalent Fourier-featured methods \citep{chandrasekhar2021tounn}, and further ensured 3D-printing capability through our introduction of connectivity constraints. Through comprehensive experiments, we explore the trade-off between structural mechanical performance and aesthetics and demonstrated the structure stylization expressiveness and controllability in various applications.

\begin{figure*}[t]
    \centering
        \includegraphics[width=\linewidth]{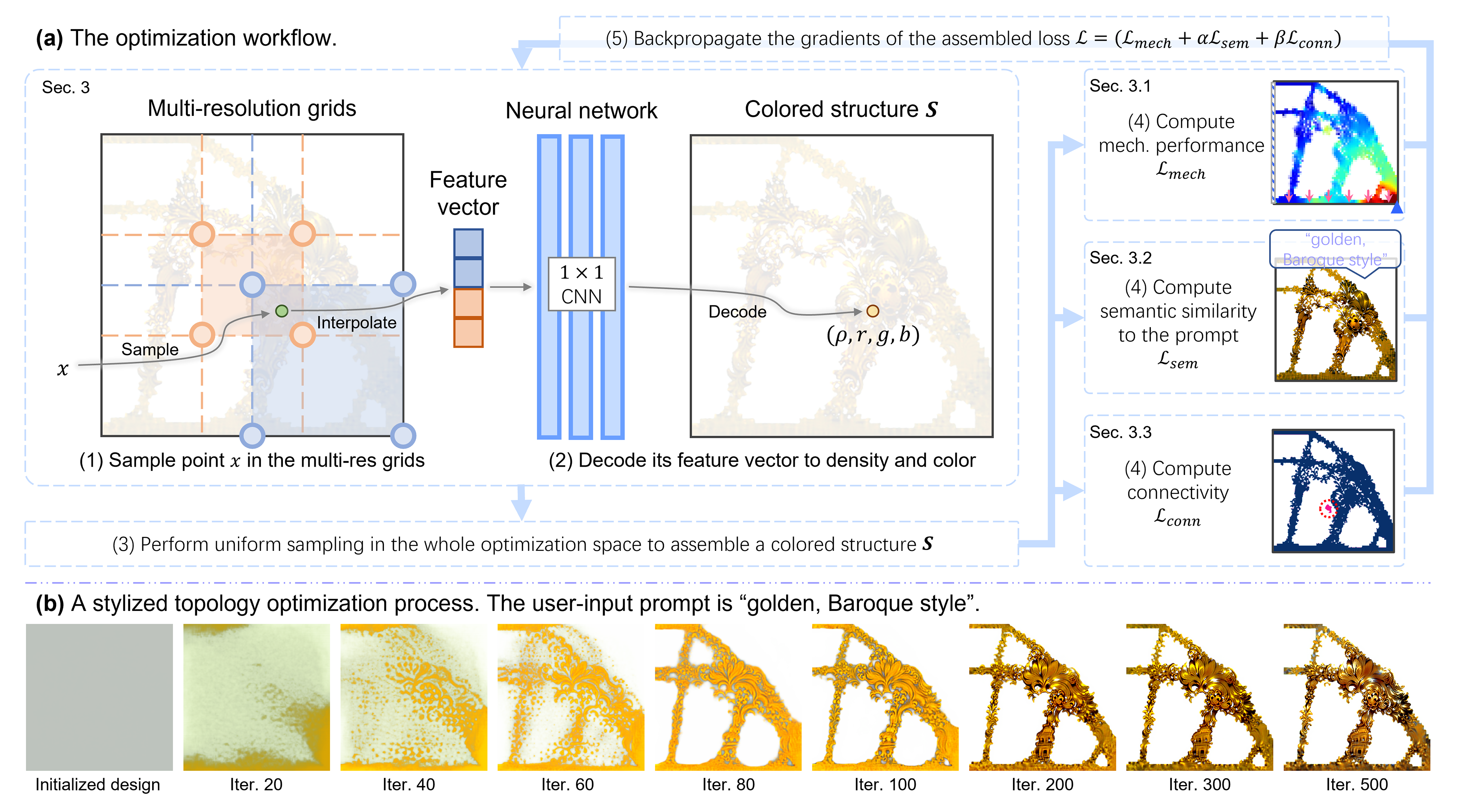}
    \caption{The system overview. (a) The overall optimization workflow. First, multi-resolution hashed grids are used to store and interpolate the coordinate-dependent features of a structure $\textit{\textbf{S}}$. Second, these features are decoded to the density $\rho$ and colors $\textit{r}, \textit{g}, \textit{b}$ using a tiny neural network. Third, through uniform sampling in the grids, a colored structure $\textit{\textbf{S}}$ is obtained. Fourth, we compute the structure's mechanical performance, semantic similarity to the user-input stylization prompt, and connectivity. Finally, these optimization objects are assembled as a single loss $\mathcal{L}$, and its gradients are backpropagated to both the grids and neural network to update the structure representation. Therefore, the three optimization objects are simultaneously optimized during each iteration. (b) The optimization process of a Bridge structure with a stylization prompt ``\textit{golden, Baroque style}". The optimization starts from a randomly initialized structure, and converges to a stable topology taking around $100$ iterations.}
    \label{fig: overview}
\end{figure*}

\begin{figure*}[t] 
    \centering
        \includegraphics[width=\linewidth]{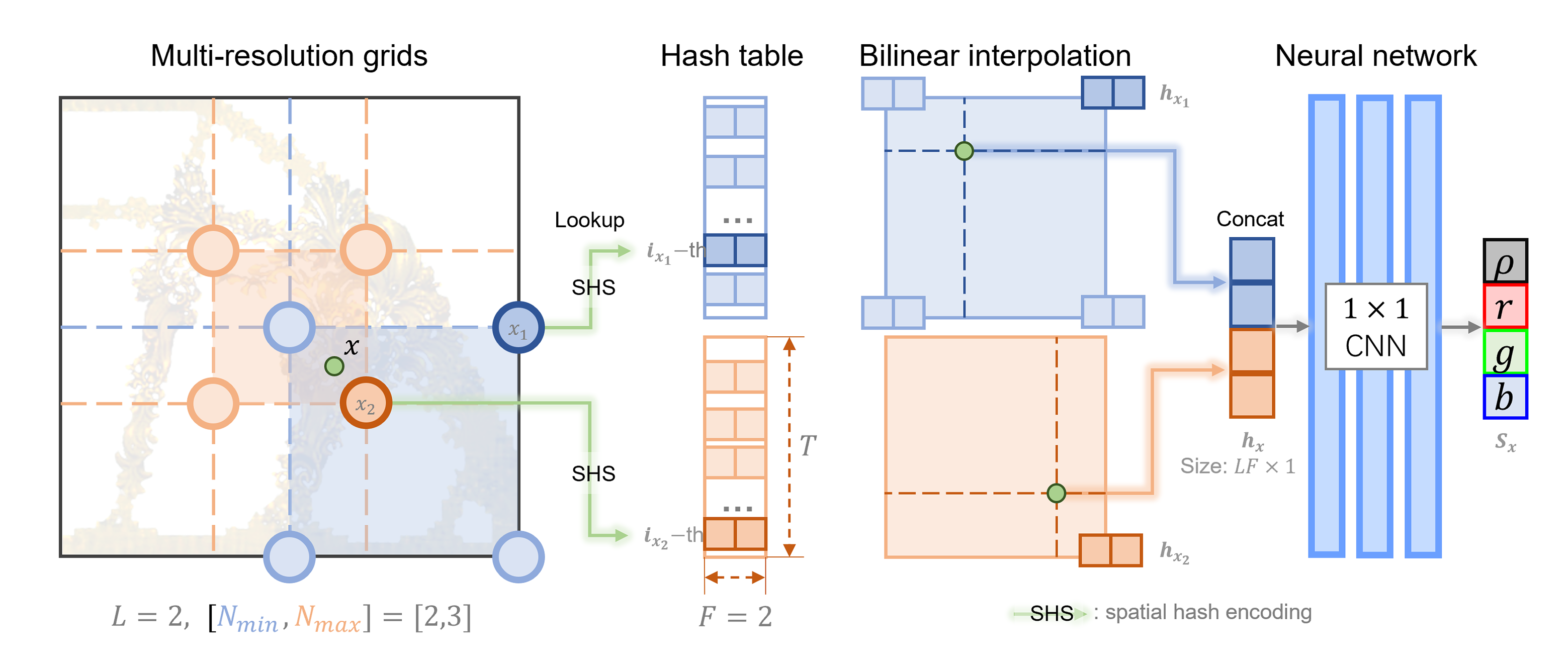}
    \caption{The Multi-resolution hash encoding process, where an input coordinate $\textit{\textbf{x}}$ is encoded to a four-dimensional output $\textit{\textbf{S}}_x$. Initially, we encode the corner vertices of the element containing $\textit{\textbf{x}}$, and employ the encoded values as indices to look up in the hash table, thus obtaining the feature vectors of the corner vertices. Subsequently, we obtain the feature vector at $\textit{\textbf{x}}$ via bilinear interpolation of the feature vectors of corner vertices. Finally, we concatenate the feature vectors of $\textit{\textbf{x}}$ at $L$ layers and feedforward through a neural network to attain the output $\textit{\textbf{S}}_x$.}
    \label{fig: ngp}
\end{figure*}

\section{Proposed Method}\label{sec3}
In this study, we present a problem formulation as a multi-objective optimization scenario. Initially, we adopt a hash-encoded neural network as a means to implicitly represent a structure $\textit{\textbf{S}}$. Subsequently, we concurrently assess the mechanical performance, aesthetic style, and connectivity aspects of $\textit{\textbf{S}}$. Finally, we update the structure $\textit{\textbf{S}}$ utilizing the gradients associated with these three objectives. The complete optimization workflow is shown in Fig. \ref{fig: overview}.

In the three subsections of Sec. \ref{sec3}, we provide a sequential account of the methodologies employed to compute each objective function as outlined below:
\begin{itemize}
\item Sec. \ref{subsec3.1}: Compute the mechanical performance ${\mathcal{L}}_{mech}$ of the structure through density-based topology optimization.
\item Sec. \ref{subsec3.2}: Compute the aesthetic style ${\mathcal{L}}_{sem}$ of the structure by the image-text neural network CLIP.
\item Sec. \ref{subsec3.3}: Compute the connectivity ${\mathcal{L}}_{conn}$ of the structure by applying the Connected Component Labeling algorithm.
\end{itemize}

The integration of the three objects is accomplished through the utilization of the penalty method, a technique that converts a constrained optimization problem into an unconstrained form, as shown in Eq. \ref{eq: loss}. Here, the penalty factors $\alpha$ and $\beta$ are assigned to govern the semantic and connectivity losses of the structure, respectively.
\begin{equation}
\label{eq: loss}
    \mathcal{L} = {\mathcal{L}}_{mech} + \alpha{\mathcal{L}}_{sem} + \beta{\mathcal{L}}_{conn}
\end{equation}

We employ the Adam optimizer \citep{kingma2014adam} with a decreasing learning rate to ensure convergence. The gradients of the loss function $\mathcal{L}$ are back-propagated to the hash-encoded network, encompassing both the hashed grid features and the network's weights, as depicted in Fig. \ref{fig: overview}. This completes a single iteration of the structure optimization.

We contemplate the selection of structural representation methods from the following perspectives. Firstly, with regard to the explicit and implicit depiction of the structure, we have opted for the latter in order to acquire a more adaptable design space. The utilization of implicit representation allows for the interpolation of the structure to higher resolutions. Moreover, by employing different resolutions for computing the objective functions (e.g., conducting FEM analysis at low resolution while optimizing style at high resolution), we can enhance computational efficiency.

Secondly, among the various implicit representation methods, we have selected neural networks to approximate the implicit representation of the structure. This choice enables us to achieve a higher degree of structural expressiveness. Conversely, alternative implicit methods \citep{wein2020review} like level sets \citep{wang2003level} or moving morphable components \citep{zhang2016new} tend to simplify the structural representation by assuming the structure is composed of basic elements. Such simplifications conflict with our objective of ensuring a rich and expressive representation of the structure's appearance.

Lastly, in our approach of utilizing neural networks to implicitly represent the structure, we have adopted a hybrid representation technique, namely the \textit{instant neural graphics primitives} \citep{muller2022instant}. Specifically, we store the spatial features of the structure in multi-resolution hashed grids, which are subsequently decoded into color and density using a neural network. Within this framework, the neural network is exclusively responsible for feature decoding and does not need to store the structural features within its network weights, as seen in previous works \citep{chandrasekhar2021tounn}. Therefore, we can employ a compact neural network for this purpose. This approach significantly enhances the convergence speed of the optimization process compared to previous studies and effectively preserves high-frequency details in the structure.

In this framework, the multi-resolution hashed grids encode the input coordinate $\textit{\textbf{x}}$ into a feature vector $\textit{\textbf{h}}_\textit{\textbf{x}}$, while the neural network $\Phi$ decodes $\textit{\textbf{h}}_\textit{\textbf{x}}$ into the structural parameters $ \textit{\textbf{S}}_{\textit{\textbf{x}}}$, as Eq. \ref{eq: coordinate enc}.
\begin{equation}
\label{eq: coordinate enc}
    \textit{\textbf{S}}_{\textit{\textbf{x}}} = \Phi(\textit{\textbf{h}}_\textit{\textbf{x}})
\end{equation}

The multi-resolution hashed grids consist of $\textit{L}$ layers of two-dimensional grids, each with a resolution of $N_{l}$, as shown in Eq. \ref{eq: ngp}. Here, $l$ is the layer index, $\textit{N}_{min}$ and $\textit{N}_{max}$ are the coarsest and finest layer resolution, respectively. Within each layer, the encoded coordinate $\textit{\textbf{x}}$ is looked up from an independent hash table $\theta_l$ with $\textit{T}$ entries and $\textit{F}$ dimensions. The multi-resolution hashed grids encompass a total of $\textit{L} \times \textit{T} \times \textit{F}$ parameters.
\begin{gather}
\label{eq: ngp}
    \left.\
    \begin{array}{ll}
    N_{l} = \lfloor{N_{min} \cdot b^l}\rfloor \\
    b = \textrm{exp}(\frac{\textrm{ln}N_{max}-\textrm{ln}N_{min}}{L-1}) 
    \end{array}
    \right.
\end{gather}\textbf{}

Fig. \ref{fig: ngp} depicts the encoding process from the input coordinate $\textit{\textbf{x}}$ to output $\textit{\textbf{S}}_x$ in a multi-resolution grid, with its two layers illustrated in blue and orange. Consider the encoding process within the $l$-th layer of the grids: firstly, we find the four corner vertices coordinates $\textit{\textbf{x}}_l$ around the input point coordinate $\textit{\textbf{x}}$. Next, $\textit{\textbf{x}}_l$ are scaled by $N_l$ and rounded down, then subjected to spatial hash encoding function \citep{hamming1952mathematical} to obtain integer indices as shown in Eq. \ref{eq: hash enc}.
\begin{equation}
\label{eq: hash enc}
    i_{x_l} = (\textit{\textbf{x}}_l \bigoplus \boldsymbol{\pi}) \bmod T
\end{equation}
Where $\bigoplus$ is a bit-wise XOR operation, and $\boldsymbol{\pi}=[1, 2654435761]$ are large prime numbers for better cache coherence.

Secondly, we perform a look-up in the $l$-th layer hash table $\theta_l$ with the indices ${i_{\textit{\textbf{x}}_l}}$, to obtain the corresponding $F$-dimensional feature vectors $\textit{\textbf{h}}_{\textit{\textbf{x}}_l}$ of the corner vertices, as shown in Eq. \ref{eq: look up}. $[\quad]$ denotes indexing.
\begin{equation}
\label{eq: look up}
    \textit{\textbf{h}}_{\textit{\textbf{x}}_l} = \theta_{l}[{i_{\textit{\textbf{x}}_l}}]
\end{equation}

Third, we bilinear-interpolate the feature vectors of corner vertices back to the input coordinate $\textit{\textbf{x}}$, as $\textit{\textbf{h}}_\textit{\textbf{x}}$.

Fourth, we repeat the above steps for all $L$ layers, then concatenate the feature vectors at all layers into an $LF \times 1$ vector. A neural network, $\Phi$, finally decodes it into the output $\textit{\textbf{S}}_x$. We utilize a two-layer convolutional neural network (CNN) with a kernel size of $1 \times 1$ to reduce network parameters.

Upon confirming the sizes of the multi-resolution grids and the network, the count of design variables remains consistent during computations at any resolution. This constancy stems from the fact that, for any given sampling position in each layer of the hashed grids, the feature value is obtained through interpolation from the neighboring grids. In the case of 2D, this entails bilinear interpolation using the surrounding four grid feature values, while in 3D, trilinear interpolation employs the surrounding eight grid feature values. As a result, the network serves solely for decoding purposes and exerts negligible influence on the structural representation, so we can maintain a constant size for the convolutional neural network (CNN) throughout computations across different resolutions. Ultimately, the number of design variables exhibits a linear relationship with the size of each layer $T$ in the multi-resolution grids and positively correlates with the number of layers $L$. In other words, $L, T, F$ determine the number of design variables besides the $1 \times 1$ CNN network. Within the specified parameter selection outlined in Tbl \ref{tbl1}, the hash-encoded network necessitates a storage capacity of $67.1$MB per structure.

The core advantage of this methodology lies in its low resource consumption and fast convergence, which is attributed to its usage of a tiny neural network. Methods like the Fourier-featured network \citep{tancik2020fourier, sitzmann2020implicit}, which is widely employed in implicit neural representation topology optimization, necessitates a much larger neural network, such as a multi-layer perceptron, to store structural information, resulting in higher memory usage, larger storage, computational burden, and difficulties in convergence. In practice, the use of multi-resolution hashed grids has been shown to reduce the iterations from hours using the Fourier-featured network to seconds in the applications like gigapixel image fitting.

\subsection{Topology optimization}\label{subsec3.1}
\begin{figure}[t]
    \centering
        \includegraphics[width=\linewidth]{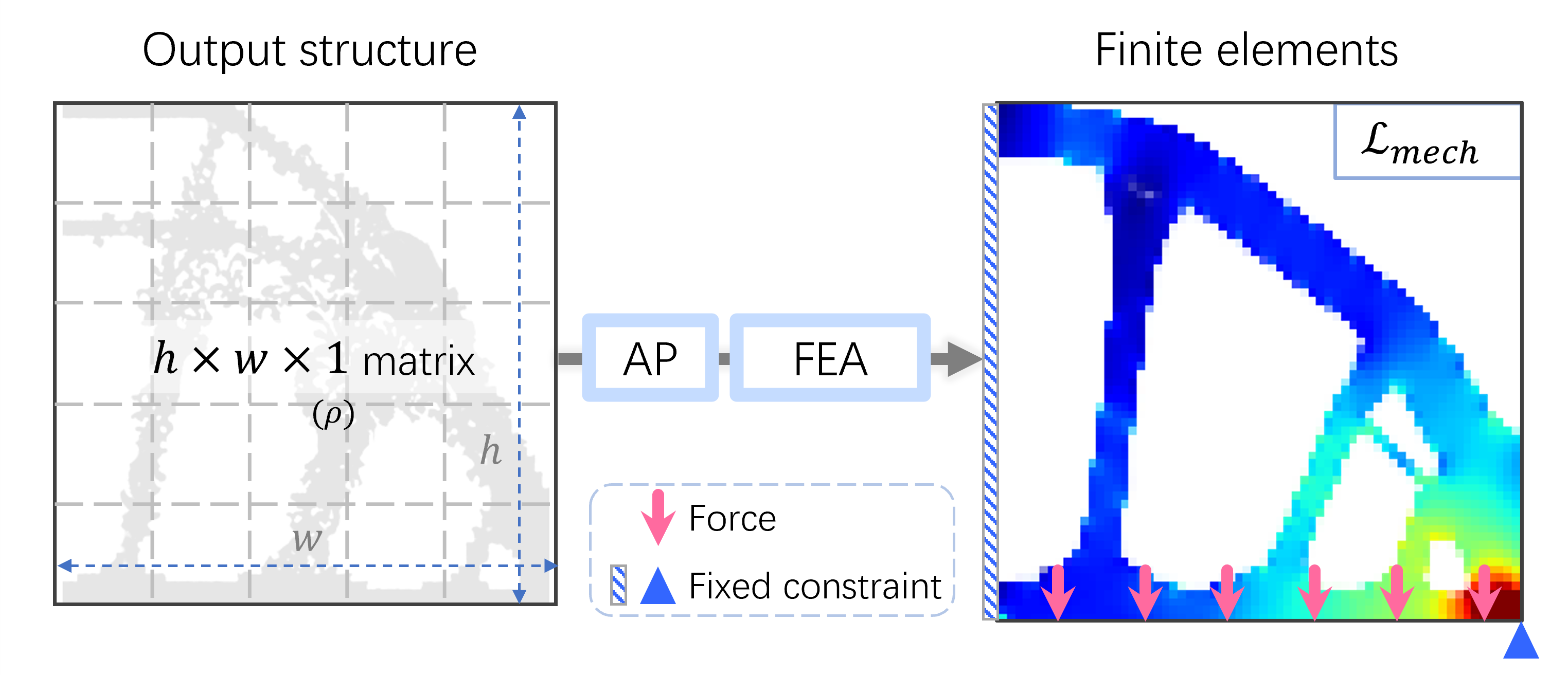}
    \caption{Structure topology optimization. Within, we perform an average pooling (AP) on the structure $\textbf{\textit{S}}$ and then conduct a finite element analysis (FEA).}
    \label{fig: to}
\end{figure}

A typical procedure for topology optimization entails conducting gradient descent on the mechanical performance of a structure to iteratively refine its volumetric representation \citep{sigmund200199}. In our study, we introduce an additional preprocessing step involving the application of average pooling (AP) to the structural density. Then, a conventional topology optimization process is performed using the finite element method (FEA). The overall workflow is shown in Fig. \ref{fig: to}.

The purpose of the average pooling is to provide more optimization space for structural stylization and alleviate the computational burden of obtaining high-resolution solutions. Specifically, following the principles of multi-resolution topology optimization, we perform topology optimization on the down-sampled grid $\textit{\textbf{S}}_{ap}$ obtained through pooling, while conducting style optimization on the original structure $\textit{\textbf{S}}$ output by the network. This approach relaxes the constraints imposed by the gradients of mechanical performance on structural details. We empirically set the kernel size and stride of the average pooling to $4 \times 4$.

In this study, compliance minimization \citep{bruggi2012topology} was adopted as the objective of the topology optimization, with the aim of achieving optimal rigidity while minimizing the weight of the structure, as shown in Eq. \ref{eq: toobject}.
\begin{gather}
\label{eq: toobject}
    \left.\
    \begin{array}{ll}
    \underset{\textbf{\textit{S}}}{\mathrm{argmin}}\, C(\textbf{\textit{S}}) \\
        s.t. \left.\
    \begin{array}{ll}
    {V(\textbf{\textit{S}})}/V_0 \le \delta
    \end{array}
    \right.
    \end{array}
    \right.
\end{gather}
Wherein, $C$ is the compliance of the structure, which reflects its deformation energy under external forces; $V$ is the volume of the structure; $V_0$ is the volume of the entire optimization space, i.e. the volume of all-filled grids; and $\delta \in (0, 1)$ denotes the user-specified objective volume fraction.

We employ an L2 loss to enforce a volume constraint on the structure and convert the constraint optimization problem into a single-objective optimization problem through the use of a penalization method, as demonstrated in Eq. \ref{eq: toloss}, where $\gamma$ is a fixed penalization factor.
\begin{equation}
\label{eq: toloss}
    \mathcal{L}_{mech} = C + \gamma (V / V_0 - \delta)^2
\end{equation}

The structure volume can be obtained by summing its density values $\rho$ over all elements, and its compliance $C$ can be calculated with the SIMP topological optimization method \citep{andreassen2011efficient} and finite element analysis \citep{rao2017finite}. The procedure is briefly outlined as follows. 

Firstly, the pooled structure $\textit{\textbf{S}}_{ap}$ is constructed as a finite element mesh comprising of rectangular elements, and the structure's stiffness matrix $\textbf{\textit{K}}$ and element stiffness matrix $\textbf{\textit{K}}_e$ are formulated based on the material elastic properties. They indicate the deformation of the structure under external loads. 

Secondly, we solve the structure's deformation $\textbf{\textit{U}}$ under the external force $\textbf{\textit{F}}$, according to the generalized Hook's law $\textit{\textbf{KU=F}}$. 

Finally, the compliance $C$ is calculated as $C = \boldsymbol{\rho} \textbf{\textit{U}}^T \textbf{\textit{K}} \textbf{\textit{U}}$, thus completing the calculation of $\mathcal{L}_{mech}$. Here $\boldsymbol{\rho}$ denotes the $h \times w \times 1$ density channel of the structure $\textit{\textbf{S}}_{ap}$. Readers may refer to established research \citep{andreassen2011efficient} for a detailed derivation of the topology optimization for compliance minimization task.

Upon establishing the value of $\mathcal{L}_{mech}$, we opt to utilize the Adam optimizer instead of the optimality criteria method (OC) \citep{sigmund200199} or the method of moving asymptotes (MMA) \citep{rojas2015benchmarking}, which are mathematical programming methods. The rationale behind this decision is: MMA, OC, and similar optimization methods are specifically tailored for topology optimization applications. They have demonstrated their effectiveness in enforcing tight constraints (e.g., volume constraint) during the topology optimization process. However, when the optimization objective involves a neural network, we favor the adoption of more versatile gradient-descent optimizers, which facilitate stable convergence. However, general neural network optimizer has significant limitations in that it's hard to reach the set constraints, and users have to manually decide the trade-off between optimization objects (e.g., in topology optimization cases, the optimized structures always have a larger volume than the objective volume fraction $\delta$.)

\subsection{Style optimization}\label{subsec3.2}
In order to stylize a structure based on a textual description, the CLIP model \citep{radford2021learning} is introduced. This is a neural network trained on a large corpus of text-image pairs. It converts images and texts into latent codes through corresponding encoders and learns the text-image matching relationship by minimizing the similarity between the latent features of images and texts. After training, it can be utilized for tasks such as image labeling \citep{zhou2022extract}, image highlighting \citep{decatur20223d}, and text-to-image synthesis \citep{frans2021clipdraw}. In this paper, our objective is to maximize the semantic score, i.e., to maximize the similarity between the image $\textbf{\textit{I}}$ (Eq. \ref{eq: alpha pen}) of the structure (with the rho-channel treated as the alpha-channel of an image) and the prompt \textbf{\textit{P}} that describe the style of the structure, as shown in Eq. \ref{eq: semobj}.
\begin{equation}
\label{eq: semobj}
    \underset{\textbf{\textit{I}}}{\mathrm{argmax}}\, \textrm{similarity}(\textbf{\textit{I}}, \textbf{\textit{P}})
\end{equation}

\begin{figure}[t]
    \centering
        \includegraphics[width=\linewidth]{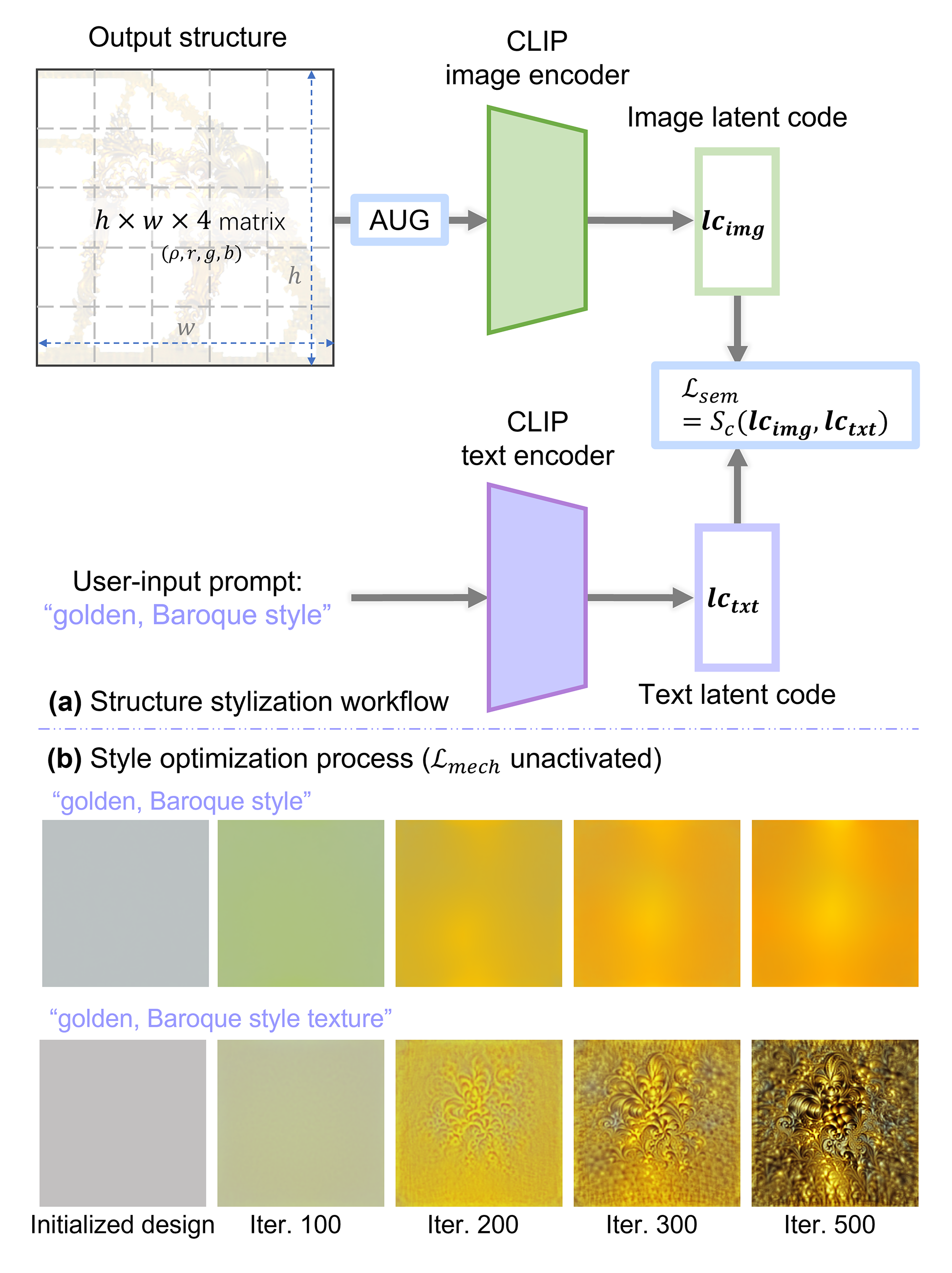}
    \caption{Structure style optimization. (a) Structure stylization workflow. Augmented image batch and descriptive prompt are input to image and text encoders of CLIP correspondingly, and the difference between the output latent codes are minimized to enforce the structure style to be consistent with the prompt. (b) Structure optimization process. We visualize the process of structures (image in this case) that are solely optimized with the semantic score $\mathcal{L}_{sem}$.}
    \label{fig: clip}
\end{figure}

The image $\textbf{\textit{I}}$ and prompt $\textbf{\textit{P}}$ are respectively encoded as 512-dimensional latent codes $lc_{img}$ and $lc_{txt}$ by the image and text encoders of CLIP, as illustrated in Fig. \ref{fig: clip} (a). The semantic similarity between them is established by utilizing cosine similarity, and negation is added to convert this semantic score into a loss function, as Eq. \ref{eq: semloss}.
\begin{equation}
\label{eq: semloss}
    \mathcal{L}_{sem} = -\textrm{cos}(\textbf{\textit{I}}, \textbf{\textit{P}})
\end{equation}

Prior to acquiring the image latent code $lc_{img}$, we perform augmentation on the image $\textbf{\textit{I}}$ in order to gain controllability of the generated results and improve the convergence. Image augmentation has been previously validated in the research of text-guided image \citep{frans2021clipdraw} and 3D shape \citep{michel2022text2mesh} generation as a means to avoid generating content with a numerically high image-text similarity but hard for humans to identify.

During each optimization iteration, we augment image $\textbf{\textit{I}}$ with a batch size of $B$. Each augmentation consists of four components: \textit{Random grayscale}, which transforms image $\textbf{\textit{I}}$ into grayscale with a specific probability, so as to encourage the system to focus more on the topology of the structure rather than merely altering textures; \textit{Random resized crop}, which randomly crops a portion of the image and resizes it to its original dimensions to focus the system on the central parts of the structure; \textit{Random affine}, which applies a random affine transform to the image to avoid generating adversarial solutions; and \textit{Random background}, which generates a random Gaussian-blurred background to avoid the system cheating the semantic score $\mathcal{L}_{sem}$ by generating textures with the same background color.

We recognize that readers may have doubts about the tendency of text-image models of stylizing the structure, i.e., that neural networks tend to optimize texture over topology to achieve higher semantic scores. We assert that this tendency can be regulated by enforcing grayscale image input to the neural network, through operations like an image alpha-channel penalty or a higher random grayscale probability. We will systematically demonstrate the controllability of the stylization in Sec. \ref{sec4}.

As for the prompt latent code $lc_{txt}$, it can be generated by feedforwarding the prompt into the text encoder. Additionally, through multiple experimental trials, we have found a correlation between the convergence speed of stylization and the choice of prompts. Generally, prompts that encompass a greater level of detail and incorporate additional semantic constraints lead to faster convergence. For instance, as depicted in Fig. \ref{fig: clip} (b), the prompt ``\textit{golden, Baroque style}" only achieves a blurry golden image after 500 iterations. However, when the prompt is completed to ``\textit{golden, Baroque style texture}", more intricate details have been obtained.

In conclusion, we adopt a trained, fixed-parameter CLIP model (\textit{ViT-B/32}) to infer the image and text latent codes, and minimize their difference to encourage the structure stylization that is semantically consistent with the text description.

\begin{figure}[b]
    \centering
        \includegraphics[width=\linewidth]{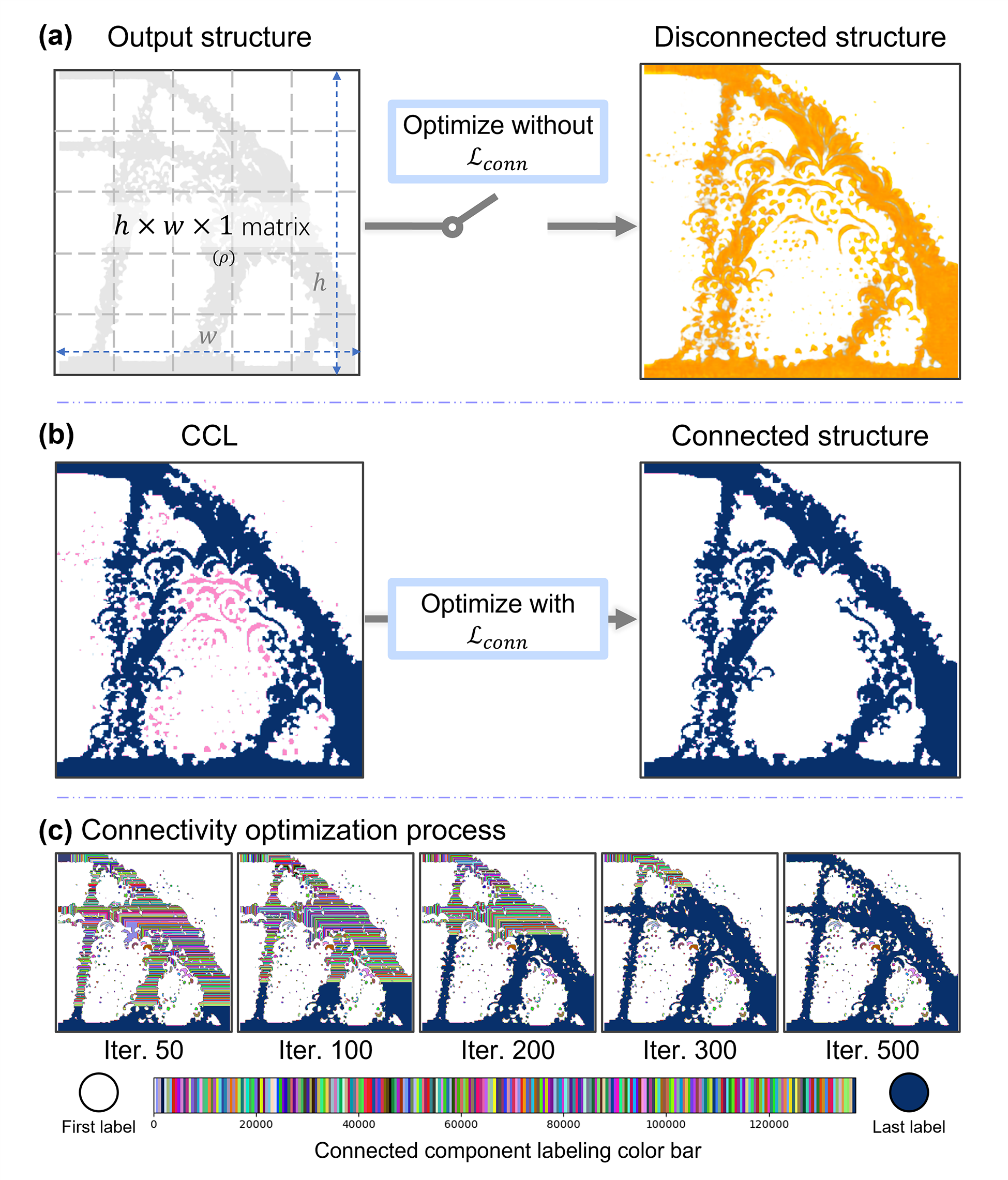}
    \caption{Structure connectivity optimization. (a) A structure could be disconnected without connectivity constraints. (b) Through applying connected component labeling, we successfully divide the disconnected parts marked in pink, and enforce them to be of zero density to optimize structure connectivity. (c) Connectivity optimization process, where colors indicate the component labels.}
    \label{fig: conn}
\end{figure}

\subsection{Connectivity optimization}\label{subsec3.3}
In the preceding two sections, we performed topology optimization on an averaged-pooled structure and subjected it to text-guided stylization. These processes may result in the presence of disconnected parts, which are meaningless in actual fabrication. Thus, we aim to introduce constraints to ensure the generated structure is integral, meaning the structural density of the disconnected parts $\rho_{d}$ should be zero. Note that the proposed connectivity constraint shares a similar motivation to the perimeter constraint, which indirectly suppresses the checkerboard pattern by minimizing the perimeter of internal boundaries of the material distribution \citep{borrvall2001topology}.

It is important to emphasize that while compliance optimization also promotes connected structures, a separate connectivity optimization step is necessary. This requirement arises due to the nature of density-based topology optimization, where a minimum density value (e.g., 1e-3) is assigned to each element in the structure to prevent numerical instabilities during convergence, rather than setting it to zero. Consequently, without connectivity optimization, the CLIP network may generate floating decorations in non-load-bearing regions of the structure in order to maximize appearance scores while incurring minimal penalties. Through empirical investigation, we have observed that this phenomenon occurs in the absence of connectivity constraints.

Therefore, we employed connected component labeling \citep{he2017connected} to identify disconnected regions within the structure and used them as a mask to construct the loss function $\mathcal{L}_{conn}$, so as to enforce the density value of disconnected parts $\rho_d$ to be zero, as shown in Eq. \ref{eq: connloss}.
\begin{equation}
\label{eq: connloss}
    \mathcal{L}_{conn} = \lvert\ \rho_{d} \rvert
\end{equation}

The process is depicted in Fig. \ref{fig: conn}. Firstly, the structure density $\rho$ was thresholded into binary values of 0 and 1, with a threshold of $0.1$, and the portion with $\rho=1$ was extracted as the mask $M$. Secondly, a labeling matrix $\textit{\textbf{Q}}$ of the same size as the structure ($h \times w \times 1$) was initialized with element values ranging from $[1, hw]$. The non-masked portion of the labeling matrix was then set to zero, i.e. $\textit{\textbf{Q}}[\sim \textit{\textbf{M}}]=0$, and the matrix was iteratively subjected to max-pooling (kernel size = 3, stride = 1, padding = 1) to obtain the labeling of the connected regions within the structure. Finally, using the volume fraction $\delta$ as a threshold, we designated regions with a label corresponding to fewer than $\delta hw$ elements as disconnected, and obtained the indices $d$ of these elements, thus making the density value of disconnected parts to be zero, $\rho_d$, in the structure.

Essentially, users need to specify one parameter for the connectivity constraint, which is the number of iterations for performing the connected component labeling (CCL). Let's assume we are performing CCL on a structure with a resolution of $h\times w$. In the worst case, where the entire structure resembles a checkerboard pattern, it would necessitate $h\times w$ iterations of CCL. In practice, due to the prevalence of extensive connected regions within optimized designs, we recommend users empirically reduce the iteration number to $0.5hw$ to expedite computation.

\section{Experiments}\label{sec4}
\textit{Experiments overview.} During the experiments, we focus on the mechanical performance of the structure (Sec. \ref{subsec4.1}), the controllability and visual quality of text-guided generation (Sec. \ref{subsec4.2}), and the connectivity of the structure (Sec. \ref{subsec4.3}). Note the results presented in Sec. \ref{sec4} are all based on 2D structures or their sketch-ups for easier understanding, and we introduce the extension to 3D structures in Sec. \ref{sec5}.

\begin{table}[b]
\begin{center}
\begin{minipage}{\linewidth}
\caption{System hyperparameters.}\label{tbl1}
\resizebox{\columnwidth}{!}{\begin{tabular}{@{}lll@{}}
\toprule
Parameter & Notation & Value\\
\midrule
Layers of multi-resolution grids & $L$ & $16$ \\
Size of hash table & $T$ & $2^{19}$ \\
Feature dimensions per entry & $F$ & $2$ \\
Grid max/min resolution & $N_{min}, N_{max}$ & $8, 256$ \\
CNN layers &  & $2$ \\
CNN channels &  & $64$ \\
Optimization iterations &  & $500$ \\
CCL iterations &  & $2000$ \\
Loss function penalty factors & $\alpha,\beta,\gamma$ & $9e3, 1, 3e3$ \\
Image batch size & $b$ & $16$ \\
% Random grayscale probability &  & $10\%$ \\
\botrule
\end{tabular}}
\end{minipage}
\end{center}
\end{table}

\begin{figure*}[t]
    \centering
        \includegraphics[width=\linewidth]{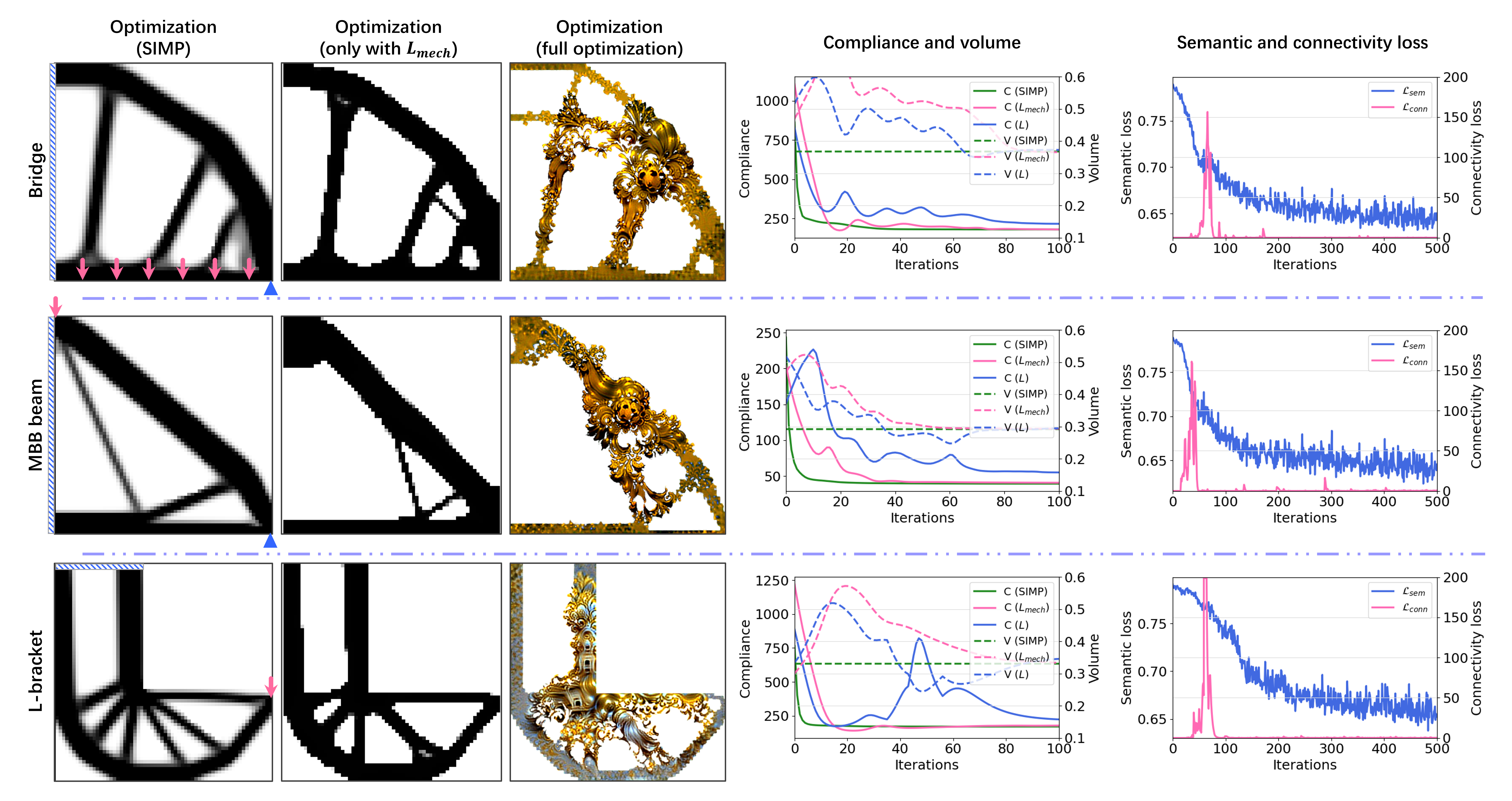}
    \caption{A comparison of the optimization performance among the traditional SIMP method, our method with only $L_{mech}$ activated, and our method with the full loss term $L$. The comparison is performed under three tasks. In the right figures, $C$ denotes the compliance and $V$ denotes the volume.}
    \label{fig: mechperform}
\end{figure*}

\begin{figure*}[t]
    \centering
        \includegraphics[width=\linewidth]{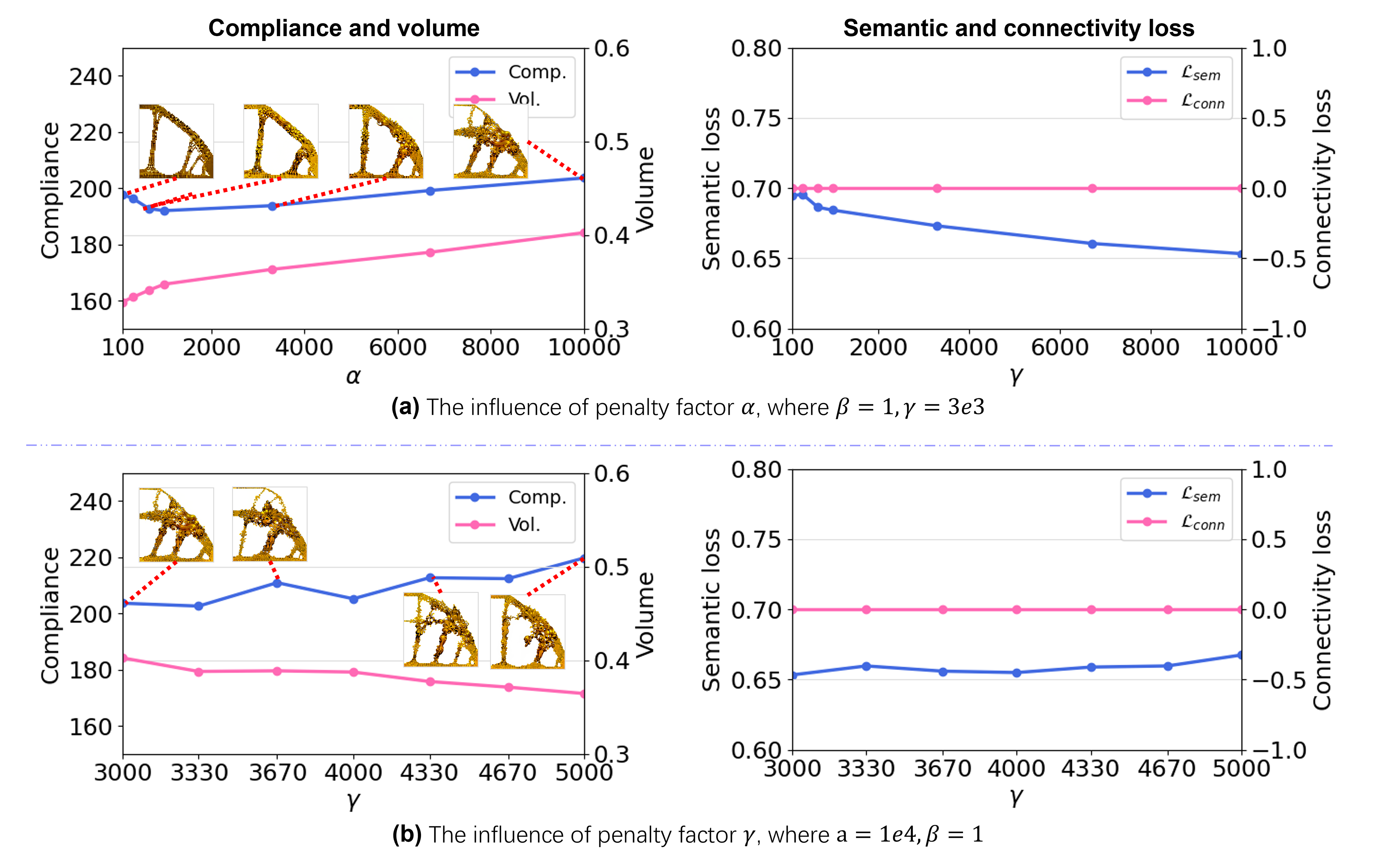}
    \caption{The impact of penalty factors on optimization results is examined by observing the optimized Bridge structure under different combinations of penalty factors. Specifically, the structure optimizations are conducted at the locations indicated by the dots in the plots, and part of the results are demonstrated. The results are pointed to their corresponding dots by red dashed lines. (a) The influence of the penalty factor $\alpha$ while keeping $\beta$ and $\gamma$ fixed. (b) The influence of the penalty factor $\gamma$ while keeping $\alpha$ and $\beta$ fixed.}
    \label{fig: penalty factor}
\end{figure*}

\textit{Experiment environment.} The proposed system runs on a laptop PC (CPU: Intel Core i9-13900HX, GPU: Nvidia RTX4080, OS: Windows11, Python3.9). In particular, we used the algebraic multigrid method \citep{wu2015system, BeOlSc2022} for an accelerated solving of the deformation matrix $\textit{\textbf{U}}$ of the structure during topology optimization. Most of the computations are run on the GPU, and VRAM is the determining factor of the computational scale.

\textit{Optimization parameters.} The hyperparameters and settings for all the experiments involved in this paper are shown in Tbl. \ref{tbl1}. Within, the scale of the hash table and CNN are positively correlated with the convergence rate, the high-frequency details of the structure, and computational cost. Our system is capable of achieving convergence within $500$ iterations for typical topological optimization tasks \citep{valdez2017topology} under various parameter settings. The CCL iterations are set to $2000$ to accommodate tasks at various resolutions. Both iteration numbers are conservatively set.

\begin{figure}[ht]
    \centering
        \includegraphics[width=\linewidth]{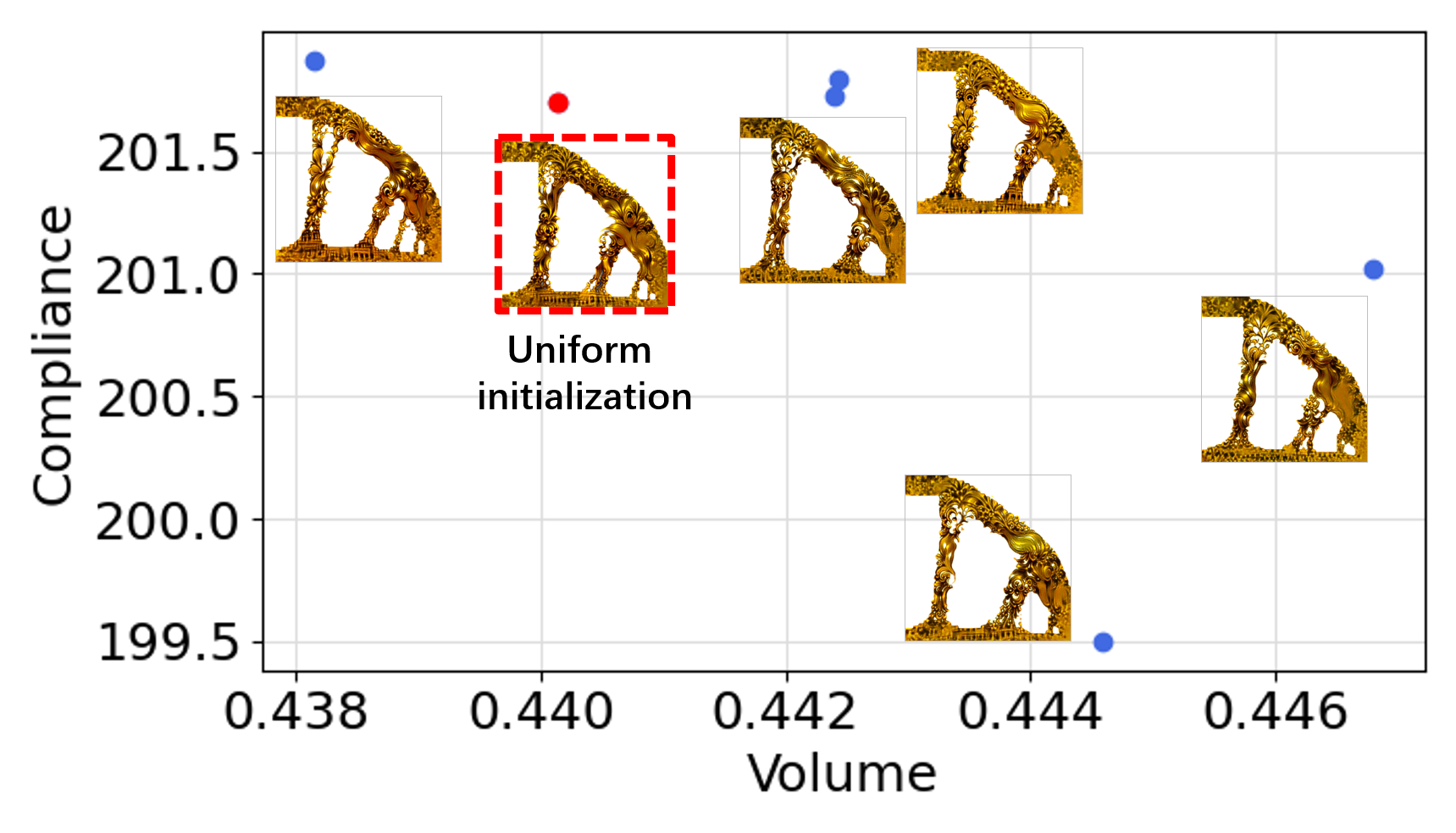}
    \caption{Randomness of our approach. The optimized structure is plotted beside its mechanical performance (the blue dots). The structure optimized with uniform design parameters initialization is marked with the red dot.}
    \label{fig: initialization}
\end{figure}

\textit{Initialization.} We randomly initialized the design parameters for allowing a certain degree of randomness in the optimized design, with the intention of enhancing the novel user experience when using our algorithm. The design variables consist of two components: feature values of multi-resolution grids, and weights of the decoding neural network (i.e., decode feature values to density and RGB values). The randomness can be disabled: First, \textit{feature values of multi-resolution grids.} We set the initial feature values of multi-resolution grids as random values between $-1^{-4}$ and $-1^{-4}$. After applying the Sigmoid activation function \citep{han1995influence}, their values will be closely centered around 0.5, resembling traditional uniform initialization. All feature values of the multi-resolution grids can be set to 0 to achieve a conventional uniform design variable distribution; Second, \textit{weights of the neural network.} Random weights of the neural network imply that under different random seeds, the network will decode the same feature value into different values. We can also fix the random seed to eliminate this randomness. The optimization solutions are influenced by the initial random design variables, as shown in Fig. \ref{fig: initialization}.

\textit{Gradients computation.} We employ the wording 'gradients' to elucidate the derivative of the loss term $L$ with respect to the design variables. Within the realm of topological optimization, ‘gradients’ is also recognized as 'sensitivity.'  To streamline the programming, we use the Autograd function of PyTorch \citep{paszke2019pytorch}, which enables automatic gradient computation (i.e., no manual derivation of the derivatives is required). The derivatives of the three terms $L_{mech}$, $L_{sem}$, and $L_{conn}$ are all computed in the same manner. The motivation for adopting automatic differentiation is its convenience. Specifically, when the system involves complex networks, manually deriving the derivatives of the loss function with respect to network parameters can be cumbersome and error-prone.

Note that new training is required for each distinct optimization task, as the hash-encoded neural network is trained to implicitly represent a single structure optimized under specific boundary conditions and stylization. In other words, each task is independent and there is no shared prior between different tasks for accelerating the training. Due to the randomly initialized design, the final solutions may have a floating performance difference around $\pm3\%$. Users may also fix the random seed to ensure the same solutions are obtained under identical input conditions.

\subsection{Validation}\label{subsec4.1}
This section presents the optimization performance of the system and verifies its core design.

\textit{Mechanical performance.} To assess the mechanical performance, we conducted an evaluation of our system by comparing the optimization performance among the traditional SIMP method (here we adopt the 165-line Python code written by Niels Aage and Villads Egede Johansen) \citep{andreassen2011efficient}, our method with only $L_{mech}$ activated, and our method with the full loss term $L$. We perform the comparison under three representative compliance minimization tasks: a Bridge, a Messerschmitt-Bölkow-Blohm (MBB) beam, and an L-bracket, as depicted in Fig. \ref{fig: mechperform}. Throughout the experiments, the inputs were specified as ``\textit{golden, Baroque style}". Finite element analysis was conducted at a resolution of $64 \times 64$ (i.e., the original $256$ resolution grids were subjected to pooling with a kernel size of $4$), while the appearance stylization optimization was performed at a resolution of $256 \times 256$ over a total of $100$ iterations.

Firstly, we compared the optimized solutions from the SIMP method and our method with only $L_{mech}$ activated, in other words, without semantic and connectivity constraints. We set the penalty factor of SIMP to be the same as of our method (p=2.0) and fine-tuned the sensitivity filter radius $r_{min}$ to 1.5 for obtaining a good optimization outcome. It turns out that our method leads to structures with sharper edges after 100 iterations. Moreover, we numerically compare the performance. As for the Bridge, MBB beam, and L-bracket optimization tasks, SIMP and our method ($L_{mech}$) yield structure with the compliance of (181.06, 181.08), (39.48, 40.80) and (169.80, 177.23), while their corresponding volumes are (0.368, 0.366), (0.293, 0.293) and (0.332, 0.333). The volume difference is due to the fact that we applied a penalty method to construct the loss, which inevitably leads to a different volume than the preset volume fraction $delta$. It turns out that our method shares a similar topology optimization performance to SIMP while featuring additional capabilities of structure resolution and pooling.

Second, we visually compared the stylized solutions to the previous ones and observed that they effectively preserve the primary load-bearing components while incorporating stylized elements around them. This observation is consistent with the mechanical performance results depicted in the ``Compliance and volume" plots of Fig. \ref{fig: mechperform}. Specifically, our method yields structures with approximately $30.27\%$ higher compliance ($19.94\%$, $39.69\%$, and $31.17\%$ higher compliance for the Bridge, MBB-beam, and L-bracket, respectively) while maintaining an enhanced aesthetic style compared to the strict compliance-minimized structures obtained through the traditional method.

\begin{figure*}[t]
    \centering
        \includegraphics[width=\linewidth]{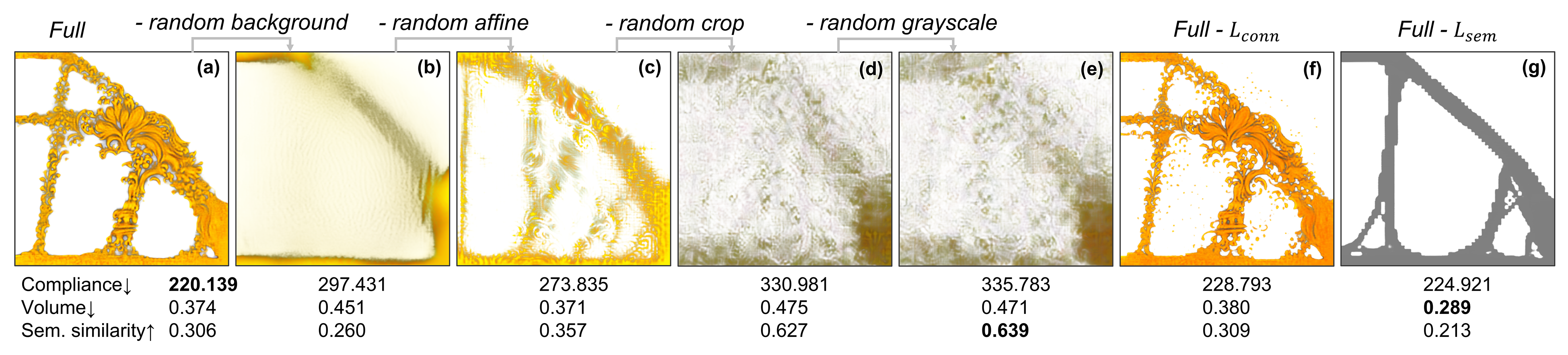}
    \caption{Ablation study of the designs in the proposed method. ($\textbf{\textit{a}} \sim \textbf{\textit{e}}$) study the image augmentation, wherein random background, random affine transform, random resized crop, and random grayscale, are successively removed. ($\textbf{\textit{f, g}}$) depict the ablation study of connectivity and stylization optimization, where '\textit{full}' represents the utilization of all proposed image augmentation and loss functions.}
    \label{fig: ablation}
\end{figure*}

Third, we observed that the proposed method tends to converge to a stable topology within $100$ iterations. Meanwhile, the colored texture of the structure takes more iterations to enrich the details. The slower convergence speed of the texture is mainly due to the nature of semantic loss $\mathcal{L}_{sem}$, as shown in the ``Semantic and connectivity loss" plots of Fig. \ref{fig: mechperform}. Here we leverage image augmentation (Fig. \ref{fig: ablation}) to relieve this issue. Image augmentation is proven effective in various research of text-guided generation \citep{michel2022text2mesh, poole2022dreamfusion, jain2022zero}, despite its random augmentation (e.g., crop, affine transform, grayscale) will lead to a noisy semantic loss. Besides, we observed that the compliance minimization process also serves as an augmentation that accelerates the overall convergence, whose effects can be visualized from the comparison between the optimization process shown in Fig. \ref{fig: overview} and Fig. \ref{fig: clip}. As for the influence of the connectivity constraint, $L_{conn}$ has substantial values only during the transition interval when the structure evolves from the initial gray density field to a connected structure (approximately within the range of 0 to 100 iterations). Subsequently, once the optimization stabilizes, we found that $L_{conn}$ always remains zero. The primary effect of the connectivity constraint lies in its capability to eliminate disconnected parts within the structure. This removal is a task challenging to achieve solely through the loss term associated with compliance minimization.

\textit{Influence of penalty factors.} The penalty factors, $\alpha$ for the semantic loss $\mathcal{L}_{sem}$, $\beta$ for the connectivity loss $\mathcal{L}_{conn}$, and $\gamma$ for the volume, collectively impact the optimization results as constituents of the loss function $\mathcal{L}$ (Eq. \ref{eq: loss}). As depicted in Fig. \ref{fig: mechperform}, the connectivity loss, $\mathcal{L}_{conn}$, remains zero for the majority of the optimization process, indicating its limited influence on the final loss function. Therefore, we focus our analysis on the effects of $\alpha$ and $\gamma$, as illustrated in Fig. \ref{fig: penalty factor}. In Fig. \ref{fig: penalty factor} (a), while keeping the volume penalty factor, $\gamma$, constant, we progressively increase $\alpha$. It is observed that $\alpha$ exhibits a negative correlation with $\mathcal{L}_{sem}$ while displaying a positive correlation with compliance and volume. Evidently, this signifies that by increasing $\alpha$, it is possible to trade off the mechanical performance of the structure for a lower semantic loss, thus achieving a higher similarity with the user-defined prompt. Similarly, in Fig. \ref{fig: penalty factor} (b), by increasing the volume penalty factor, $\gamma$, the structure's volume can be reduced, albeit at the cost of an increase in semantic loss.

It is noteworthy that the selection of $\alpha$, $\beta$, and $\gamma$ depends on various factors, including the boundary conditions of the topology optimization problem or user input prompts, introducing a degree of uncertainty. To alleviate the difficulties associated with user penalty factor selection, we propose a simple method. Upon examining the magnitudes of the different terms in the loss function, compliance is determined by the boundary conditions of the topology optimization, semantic similarity takes values between 0 and 1, while connectivity plays a minor role in the optimization process. Therefore, we neglect the impact of $\mathcal{L}_{conn}$ on the loss function $\mathcal{L}$ by setting $\beta = 1$. Additionally, we only activate $\mathcal{L}_{mech}$ to evaluate the convergence of the structure's compliance, and empirically set $\alpha$ and $\gamma$ to be one order of magnitude larger than the compliance. Consequently, the weighted terms in the loss function are of similar magnitudes. After initializing the weights using the aforementioned method, users can adjust the weights within one to two orders of magnitude to achieve personalized design requirements. Furthermore, we can generate a series of optimized solutions under different penalty factors and employ user-in-the-loop Bayesian optimization to select appropriate penalty factors.

\begin{figure}[t]
    \centering
        \includegraphics[width=\linewidth]{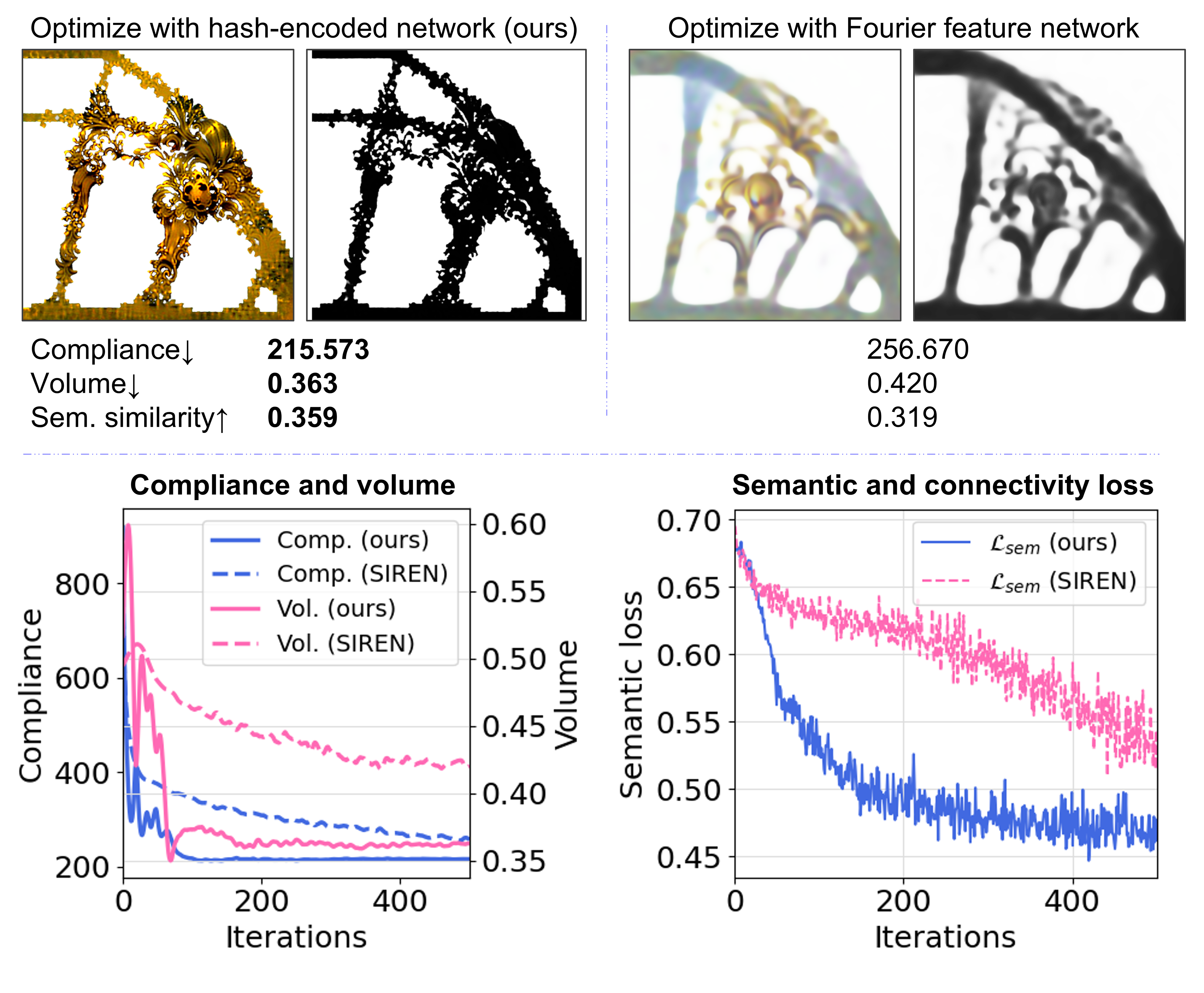}
    \caption{A comparison of the optimized results after $500$ iterations to the methods using the Fourier-featured network SIREN.}
    \label{fig: vs siren}
\end{figure}

\begin{figure}[t]
    \centering
        \includegraphics[width=\linewidth]{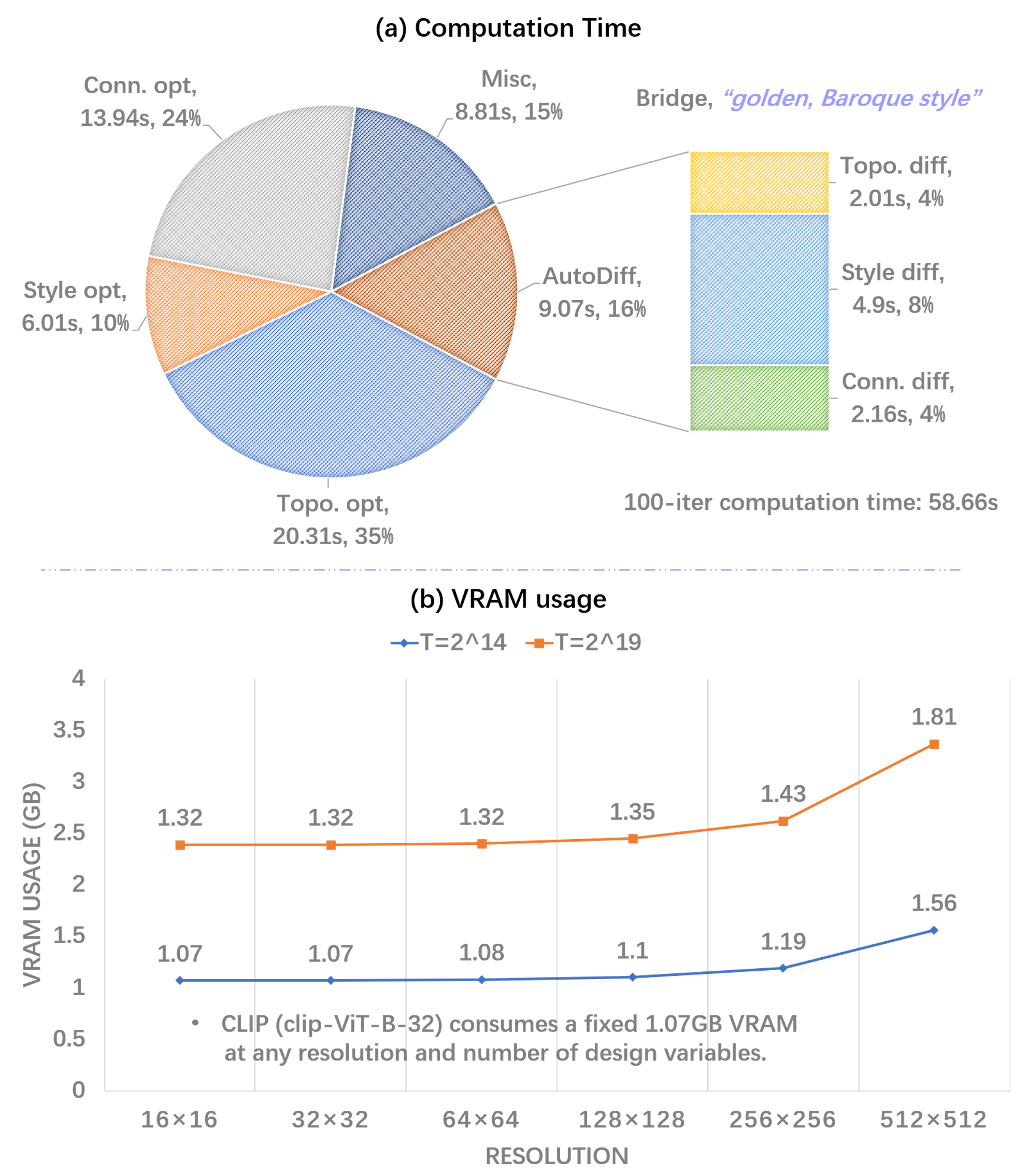}
    \caption{Computational cost. (a) The 100-iteration computation time of a bridge structure under the stylization prompt “golden, Baroque style.” (b) The VRAM usage under different resolutions and number of design variables.}
    \label{fig: computation cost}
\end{figure}

\begin{figure*}[t]
    \centering
        \includegraphics[width=\linewidth]{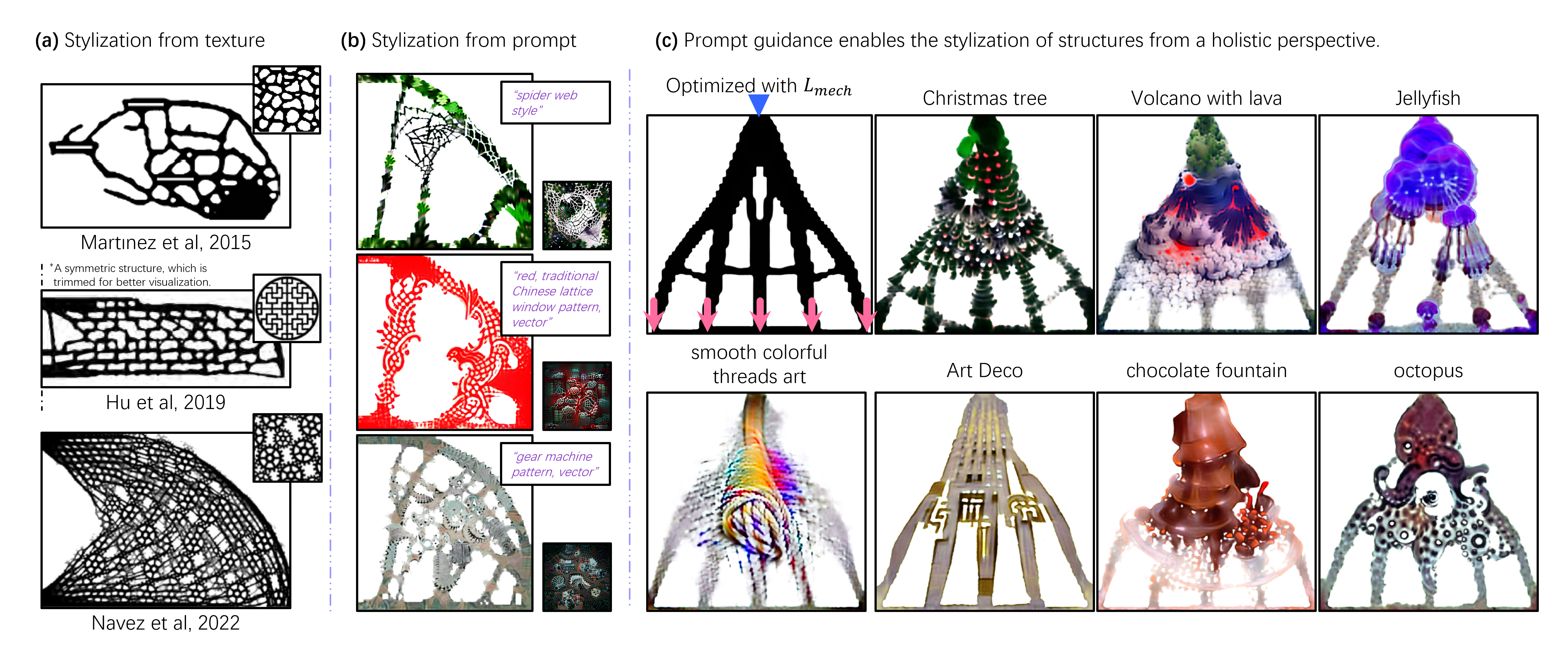}
    \caption{Visual comparison between texture-guided and prompt-guided stylization. (a) Texture-stylized results from the state of the art research. (b) Prompt-stylized results. The tiny figures in the lower right are optimized only with the semantic constraint $L_{sem}$ by our method. (c) Prompt guidance enables unique stylization.}
    \label{fig: style compare}
\end{figure*}

\textit{Convergence.} We compare the optimization convergence of our multi-resolution hash-encoded network and the Fourier-featured network \citep{chandrasekhar2021tounn} as implicit-neural-representation-based topology optimization techniques, as illustrated in Fig. \ref{fig: vs siren}. Both methods encode the input coordinates to generate the respective feature values (e.g., RGB color of an image, or density of a structure). In the comparison experiment, the hashed grids utilized a two-layer CNN, while the Fourier-featured network employed a three-layer sinusoidal activated multi-layer perceptron of 512 layer widths, with the first sinusoidal activation layer set at a frequency of $90$ to ensure the capture of adequate high-frequency structural details. Upon completion of $500$ iterations, it was found that our hashed-grids-based method obtained a more abundant representation of high-frequency structural details and better mechanical and aesthetic performance, thereby validating the enhancement in convergence speed achieved through the reduction of neural network parameters.

\textit{Computational cost.} We examine the training time and memory consumption of the optimization of a bridge with a resolution of $256 \times 256$, as shown in Fig. \ref{fig: computation cost} (a). Under 100 optimization iterations, the average iteration time was $0.587$s (i.e., a total time of $58.66$s), with the computation time for the loss functions of topology, appearance semantic score, and connectivity being $0.203$s, $0.060$s, and $0.139$s, respectively. In other words, topology optimization consumes the most computation (training) time as $35\%$, while style optimization and connectivity optimization costs $10\%$ and $24\%$. The rest training time is spent on the feedforward and backpropagation (i.e., automatic differentiation) of the neural network. The peak memory consumption under the $256 \times 256$ resolution during training was $1.43$GB, which is within the computational capability of mainstream commercial GPUs, shown in Fig. \ref{fig: computation cost} (b). Within, CLIP (clip-Vit-B-32) consumed a fixed amount of VRAM as 1.07GB. A trained network under the settings listed in Tbl. \ref{tbl1} costs $67.1$MB for storage.

\textit{Ablation study of image augmentation and loss terms.} The augmentation of structural images plays a significant role in both convergence speed and quality. Here we visualize their effects under $100$ iterations of optimization. In Fig. \ref{fig: ablation}(a$\sim$e), image augmentations were successively removed to observe their impact. The result indicated that: the presence of a random background is crucial to avoiding the generation of adversarial content. Without it, the network tends to generate textures of the same color as the background in an effort to cheat for a higher semantic loss. The random affine transforms and random resized crop ($10\%$ of the image) focus the network's attention on local regions of the structure, allowing for fine-grained updates to be made to both topology and texture. The random grayscale, which randomly converts $10\%$ of the image to grayscale, encourages the network to focus more on the structure's topology than its texture. In Fig. \ref{fig: ablation}(e), the removal of the connectivity loss terms reveals a notable inhibitory effect on disconnected parts within the structure. Lastly, the stylization loss term $\mathcal{L}_{sem}$ was removed to serve as a reference in the absence of stylization.

\textit{Comparison with texture-guided stylization}. Finally, we perform a visual comparison between texture-guided \citep{martinez2015structure, hu2019texture, navez2022topology} and our prompt-guided stylization, as shown in Fig. \ref{fig: style compare}. The results indicate that owing to the highly abstract nature of semantic representations, we can stylize the structure from a holistic perspective, which would be much more difficult for texture-guided methods.

\subsection{Stylization gallery}\label{subsec4.2}
In this section, we present the controllability of stylization and a stylization gallery.

\begin{figure}[b]
    \centering
        \includegraphics[width=\linewidth]{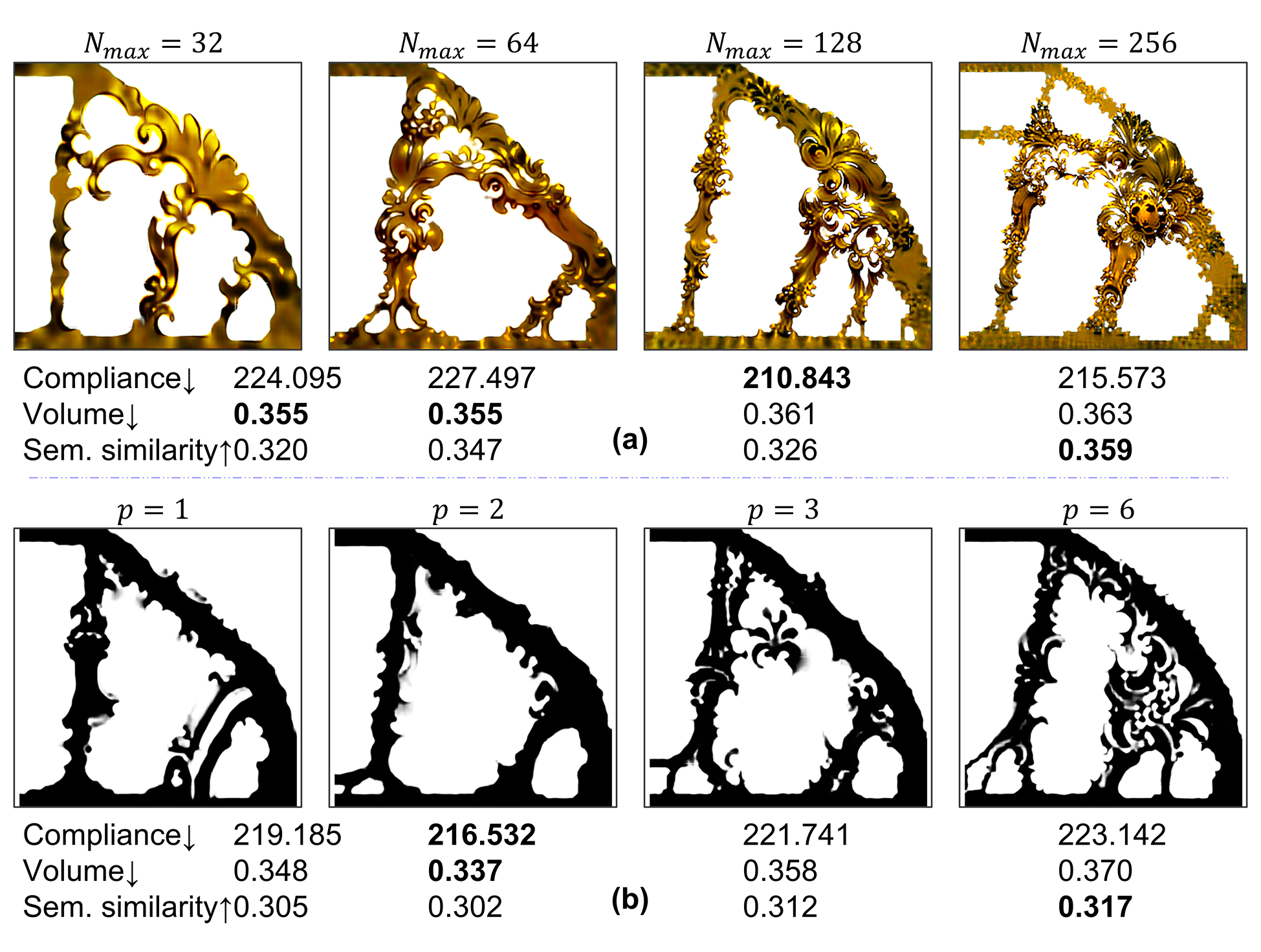}
    \caption{Stylization controllability. (a) The structural topology and texture characteristics are modulated by adjusting the maximum resolution, $N_{max}$, of the hashed grids. (b) The topology-only stylization is amplified by promoting grayscale image inputs and increasing the penalization, $p$, of the structure image alpha channel.}
    \label{fig: style control}
\end{figure}

\begin{figure*}[t]
    \centering
        \includegraphics[width=\linewidth]{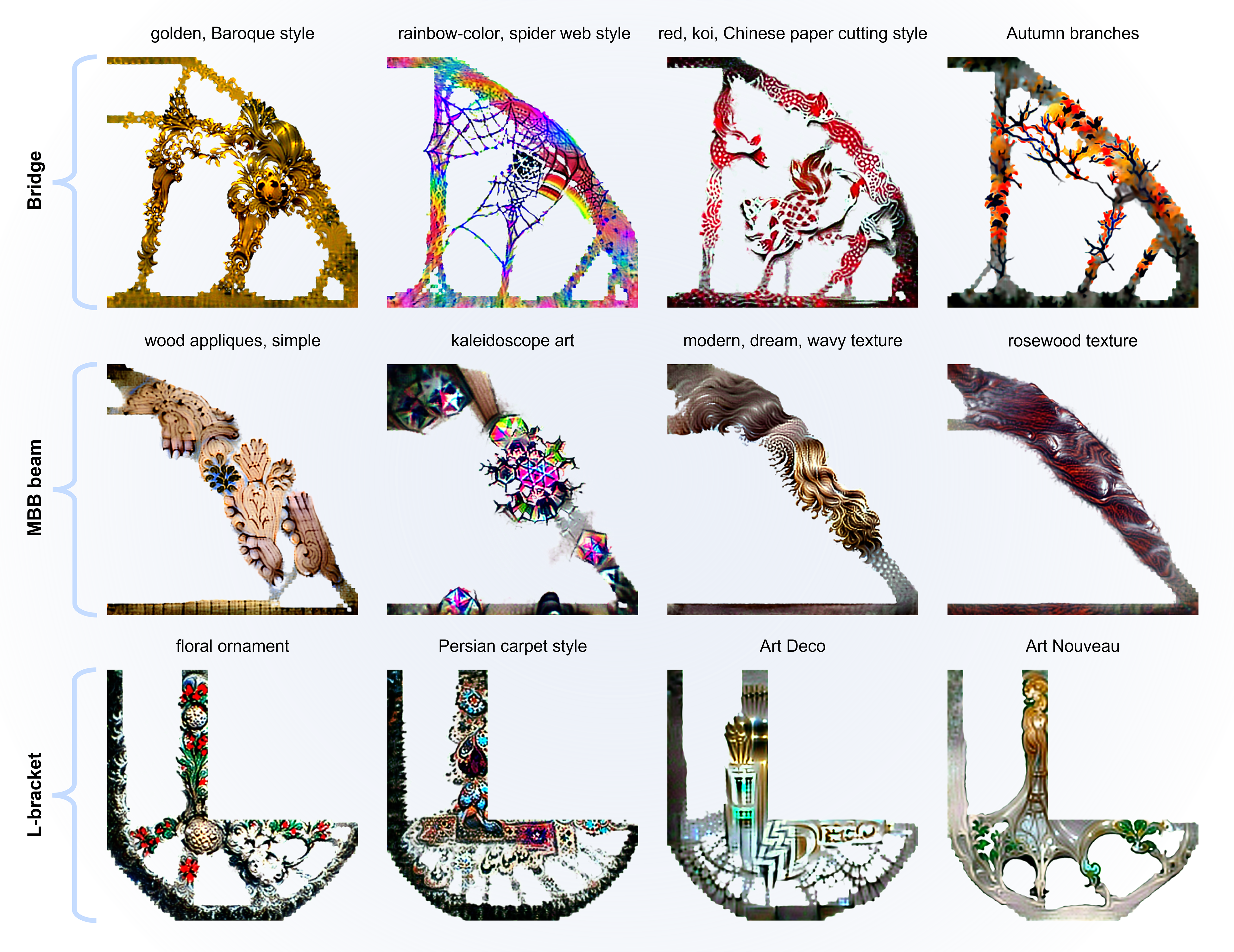}
    \caption{Stylization gallery of topology optimized structures, including (top) bridges, (middle) MBB beams, and (bottom) L-brackets. The prompts that were used for stylization are shown above each structure.}
    \label{fig: gallery}
\end{figure*}

\begin{figure*}[t]
    \centering
        \includegraphics[width=\linewidth]{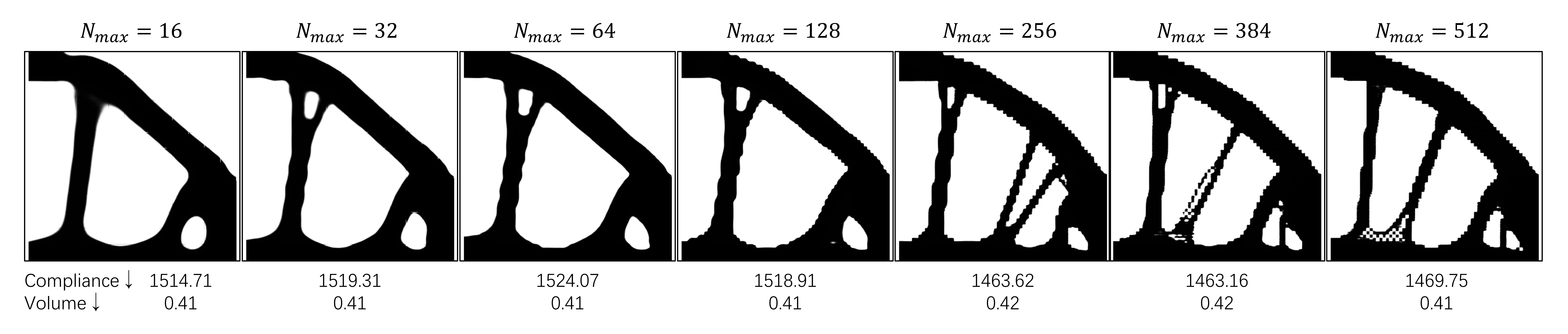}
    \caption{Length scale control of the structure through progressively tuning $N_{max}$.}
    \label{fig: length scale}
\end{figure*}

\textit{Stylization controllability.} In addition to modifying the description prompt, the structure style can also be controlled by adjusting the hyperparameters of the system. In Fig. \ref{fig: style control}(a), the control of the high-frequency details (i.e., length scale) of the structure's topology and texture is achieved by adjusting the maximum resolution $N_{max}$ of the multi-resolution hashed grids, as shown in Fig. \ref{fig: length scale}. Specifically, $N_{max}$ is negatively correlated with the length scale of the structure. The reason is that when $N_{max}$ takes a small value, the features of the structure (i.e., density and color) will be interpolated from a sparser grid, which is similar to applying a low-pass filter on the geometry and color of the structure over the design space.

Given that the topology and texture features of the structure are obtained from the bilinear interpolation of hashed grids vertices values, the maximum resolution $N_{max}$ of the grids directly determines the level of detail in the solution, and reducing $N_{max}$ can be viewed as adding a low-pass filter to the solution. The same concept is applied in the length-scale control of density-based topology optimization: filters are applied to avoid the checkerboard pattern. From a frequency domain perspective, this is equivalent to applying a low-pass filter to the density field of the structure, i.e., removing high-frequency components (i.e., checkerboard) at each optimization iteration. Consequently, this regulation will assist the user in balancing the trade-off between structural details and manufacturing difficulty or cater to a personal aesthetic sense.

In Fig. \ref{fig: style control}(b), the focus is on the system's ability to optimize the structural topology. In cases where color 3D manufacturing may be unavailable, we expect the system to still be able to express its stylization through the topology. To encourage the network to focus on the structural topology, stronger grayscale inputs are encouraged. Enhancing topological stylization involves two steps: first, the structure images are all converted to grayscale; second, the image's transparency is penalized by a factor $p$, as shown in Eq. \ref{eq: alpha pen}.
\begin{equation}
\label{eq: alpha pen}
    \textit{\textbf{I}} = \textit{\textbf{Y}} \boldsymbol{\rho}^p + \textit{\textbf{Z}} (1 – \boldsymbol{\rho}^p)
\end{equation}
where $\textit{\textbf{I}} \in \mathbb{R}^{h \times w \times 3}$ is the structural image, $\textit{\textbf{Y}} \in \mathbb{R}^{h \times w \times 3}$ is the RGB channel of the image, $\boldsymbol{\rho} \in \mathbb{R}^{h \times w \times 1}$ is the density channel of the structure (i.e., the alpha channel of the image), $\textit{\textbf{Z}} \in \mathbb{R}^{h \times w \times 3}$ is the random background, and $p$ is the penalty factor. As $p$ increases, the network's output becomes increasingly binary, thus the optimization will also become more focused on the structural topology, which is validated by the increasing semantic score. Note that we applied the same volume constraint in the optimizations presented in Fig. \ref{fig: vs siren}, Fig. \ref{fig: ablation} and Fig. \ref{fig: style control} for enabling a consistent comparison.

\textit{Stylization gallery.} We have validated the efficacy of stylization in three prototypical topology optimization cases, as shown in Fig. \ref{fig: gallery}. The results demonstrate that the system not only generates textures that align with textual description, but also stylizes the structure in one piece rather than repetitively mimicking local texture patterns (e.g., Baroque-style decorative patterns, spiderweb patterns, and branch patterns). Please take note that our proposed methodology facilitates the optimization of numerous abstract aesthetics (e.g., "\textit{wood appliques}" and "\textit{floral ornament}") that pose a challenge to be solely represented by a single exemplar, as demonstrated in Sec. \ref{subsec2.2}. Specifically, the application of periodic replication of exemplars on a structural level proves to be more suitable for stylizations encompassing periodic geometric features, such as the "\textit{Eiffel Tower}" or a "\textit{spider's web}". These approaches, however, encounters difficulties when dealing with intricate or abstract stylized objectives. Consequently, the above observation provides a justification for considering the holistic optimization of textual descriptors.

\begin{figure}[t]
    \centering
        \includegraphics[width=\linewidth]{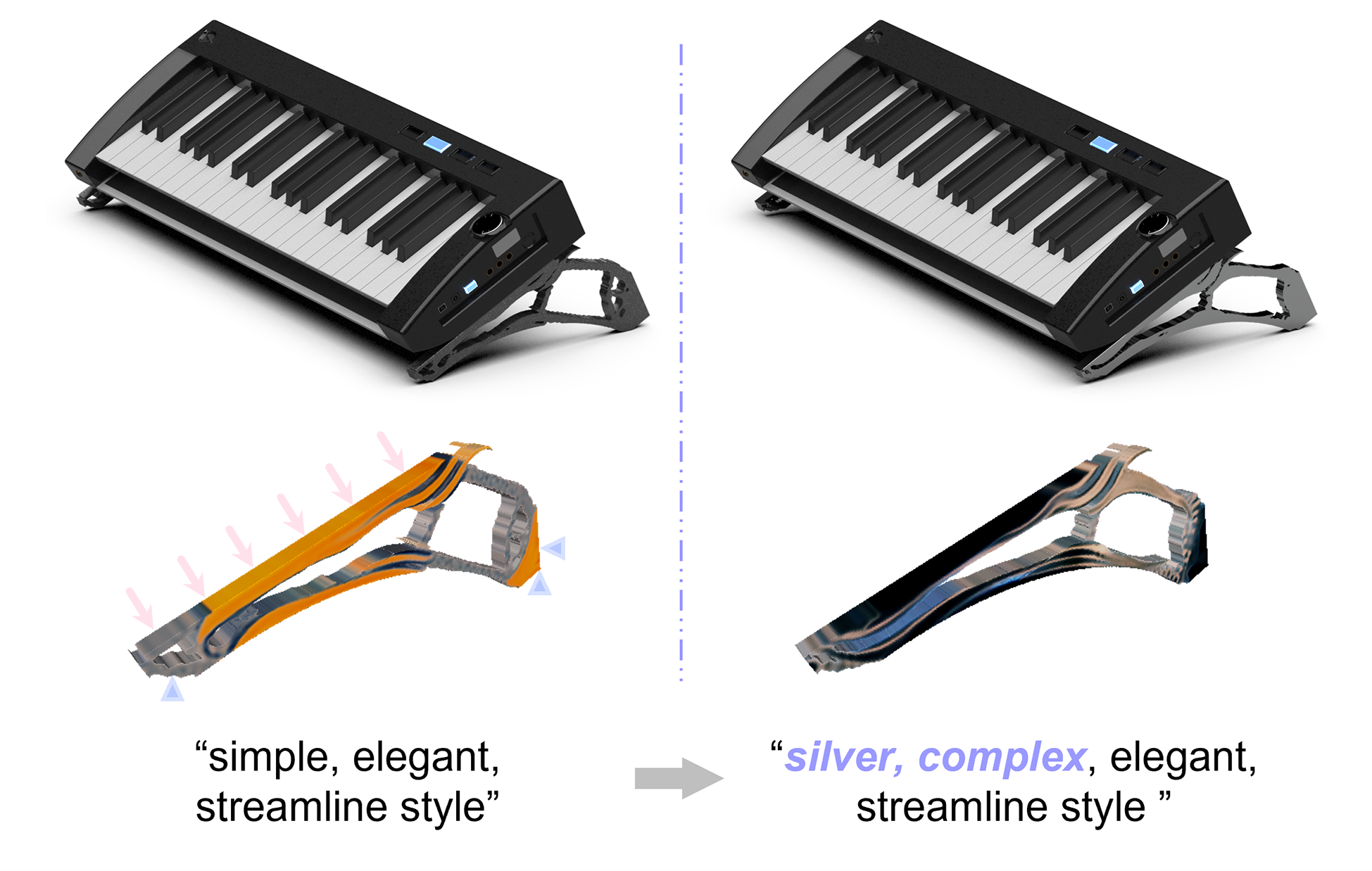}
        \caption{Editing text-based stylization of a synthesizer stand. The textual description ``streamline style" maintains the geometric feature, and the edited text ``silver, complex" leads to an appearance update.}
    \label{fig: synth demo}
\end{figure}

Moreover, users may tweak the input text to fine-tune the appearance, as shown in Fig. \ref{fig: synth demo}. While simultaneously achieving performance and global stylization in structural design remains a challenging task, we believe that the proposed system provides users with an accessible and efficient tool to rapidly iterate ideas.

\subsection{Full-color 3D Printing}\label{subsec4.3}
\begin{figure}[t]
    \centering
        \includegraphics[width=\linewidth]{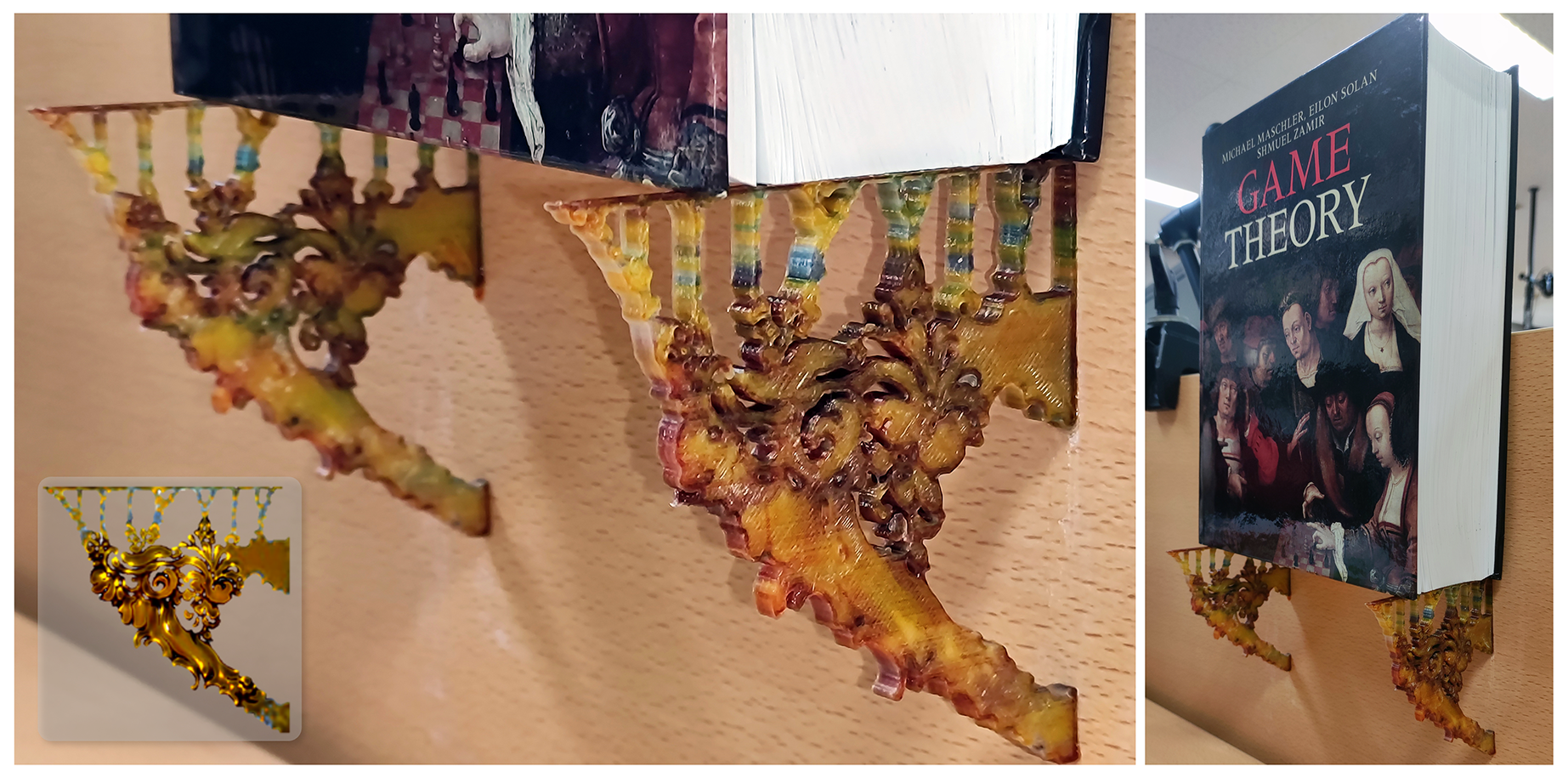}
    \caption{3D printing results of stylized topological optimized structures. The prompt was ``\textit{golden, Baroque style}". The stylized structures were 3D-printed using the \textit{XYZ da Vinci Color mini}.}
    \label{fig: 3d print}
\end{figure}

We 3D-printed sample objects generated from our method to validate the connectivity of the structures. As demonstrated in Fig. \ref{fig: 3d print}, we printed a set of topologically optimized bookshelves, which are subjected to distributed loads on the top surface and fixed on the sides. After completing the optimization, we fed refined grids into the network to obtain higher-resolution structures with smoother surface contours \citep{chandrasekhar2021tounn}. The 2D structures were then sketched and converted into a mesh in PLY format using the marching cube method \citep{lorensen1987marching}, with structural textures represented by vertex color. The experimental results attest to the significance of incorporating a connectivity loss term in the optimization process and exhibit the structural capacity of weighty external loads.

\begin{figure}[b]
    \centering
        \includegraphics[width=\linewidth]{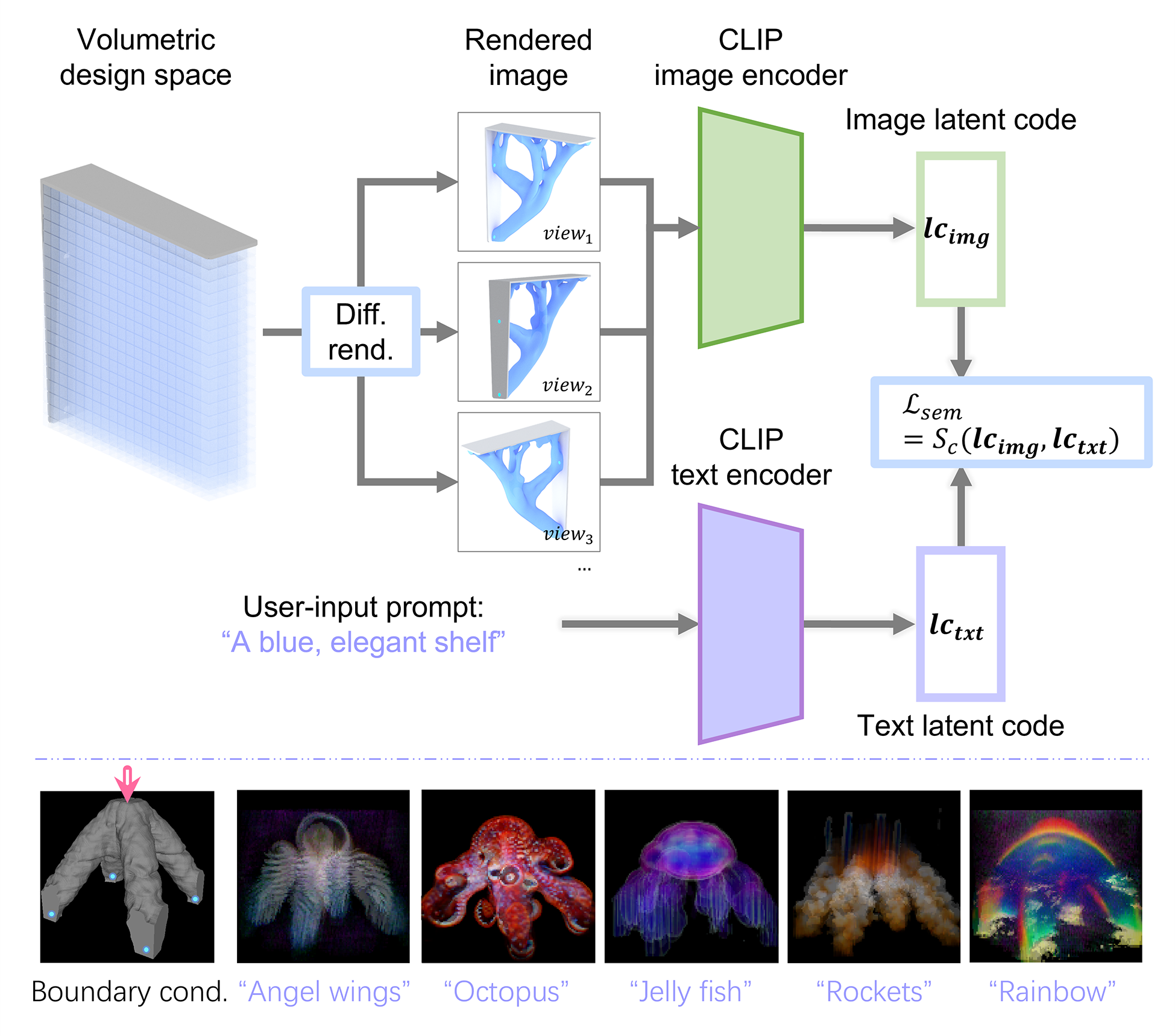}
    \caption{The optimization workflow for 3D stylized topology optimization. The main difference with 2D cases is that images are rendered from the 3D volumetric representation rather than directly obtained. Then, multi-view images are feedforward to the CLIP network for assessment of their semantic similarity to the user-input prompt. We visualize several optimized 3D structures below, labeled by their stylization prompts respectively.}
    \label{fig: 3d to}
\end{figure}

\section{Limitations and future works}\label{sec5}

The system can be extended to stylized topology optimization in 3D, as shown in Fig. \ref{fig: 3d to}. The primary difference between 3D and 2D problems lies in the fact that the structure images are obtained from the differentiable rendering $\textit{\textbf{I}} \in \mathbb{R}^{h \times w \times 3}$ of a 3D volumetric representation, which is a 3D grid with four channels of colors and density: $\textit{\textbf{S}} \in \mathbb{R}^{x \times y \times z \times 4}$. This is achieved by leveraging the neural radiance field \citep{mildenhall2021nerf}, which samples the structural features along the camera view directions within the optimization space, and accumulates these features into the pixels corresponding to each camera view, so as to render an image. The computational cost of obtaining high-quality optimized structures significantly increases with the dimensionality. Currently, our pipeline that performs topology and appearance optimization synchronously in each iteration often requires tens of minutes to produce a well-stylized 3D result.

Another major drawback of the proposed method lies in the application of penalty optimization. In this case, it is much more difficult to reach a desired volume fraction (or other pre-defined constraints) due to the nature of penalty optimization. The same situations are happening to stress and displacement-constrained optimization. This limitation is also noticeable when performing pure topology optimization tasks: both the optimization quality and the convergence speed are not as good as traditional explicit methods.  Introducing conventional optimizers like MMA into the current method would be a promising solution.

In future work, we will enhance the parameterization of the system to gain more controllability over stylization (e.g., parameterize the structure with skeletons for manual shape adjustment). Additionally, we believe that extending the system for multi-material 3D printing is also a promising avenue.

\section{Conclusions}\label{sec6}
The simultaneous pursuit of functional and aesthetical design in commercial or personalized products has long been a challenging task, requiring designers to possess a sound understanding of physics and an impeccable sense of aesthetics. We present a text-guided stylized topological optimization method, achieved through the introduction of a large-scale text-image neural network. Upon input of mechanical design requirements and a textual description of desired structure style, our system is capable of generating full-color, 3D printable solutions with stylistic tunability. 

We consider the proposed system both as a fabrication-ready design tool for DIY enthusiasts, and a backbone and source of inspiration for advanced structure stylization design using powerful and controllable generative methods \citep{zhang2023adding}. At present, there are numerous intriguing issues that remain to be investigated, such as the stylization of multi-material structures and part-aware shape parameterization \citep{hertz2022spaghetti}. We believe these developing techniques will finally aid in making topology optimization a more user-friendly automated tool, improving design efficiency and inspiring design creativity.

\section*{Statements and Declarations}
\textbf{Competing Interests:} The authors declare that they have no known competing financial interests or personal relationships that could have appeared to influence the work reported in this paper.

\noindent\textbf{Acknowledgments and Funding Information:} We are grateful to the anonymous reviewers for their suggestions and comments. This work was partially supported by the Japan Society for the Promotion of Science KAKENHI Grant Number JP19K20321 and JP20H05958, and the Japan Science and Technology Agency ACT-X Grant Number JPMJAX20AK. 

\noindent\textbf{Replication of results:} Our code for stylized topology optimization is available at https://github.com/shzzhong/TGTO.

\bibliography{sn-bibliography}% common bib file

\end{document}